\documentclass{emulateapj}
\usepackage{apjfonts}
\begin{document}

\maxdeadcycles=10000

\title{The Detailed Star Formation History in the Spheroid, Outer
Disk, and Tidal Stream of the Andromeda Galaxy\altaffilmark{1,2}}

\author{Thomas M. Brown\altaffilmark{3}, Ed Smith\altaffilmark{3}, Henry C. Ferguson\altaffilmark{3}, R. Michael Rich\altaffilmark{4}, Puragra Guhathakurta\altaffilmark{5}, Alvio Renzini\altaffilmark{6}, Allen V. Sweigart\altaffilmark{7}, \& Randy A. Kimble\altaffilmark{7}}

\submitted{Accepted for publication in The Astrophysical Journal}

\altaffiltext{1}{Based on
observations made with the NASA/ESA Hubble Space Telescope, obtained
at the Space Telescope Science Institute, which is operated by AURA,
Inc., under NASA contract NAS 5-26555. These observations are
associated with proposals 9453 and 10265.}

\altaffiltext{2}{Some of the data presented herein were obtained
at the W.M. Keck Observatory, which is operated as a scientific
partnership among the California Institute of Technology,
the University of California, and NASA.  The Observatory was
made possible by the generous financial support of the W.M. Keck 
Foundation.}

\altaffiltext{3}{Space Telescope Science Institute, 
3700 San Martin Drive, Baltimore, MD 21218;  tbrown@stsci.edu, 
ferguson@stsci.edu, edsmith@stsci.edu}

\altaffiltext{4}{Division of Astronomy, Dpt.\ of Physics \& Astronomy, 
UCLA, Los Angeles, CA 90095; rmr@astro.ucla.edu}

\altaffiltext{5}{University of California, 271 Interdisciplinary 
Sciences Building, 1156 High Street, Santa Cruz, CA 95064; raja@ucolick.org}

\altaffiltext{6}{Osservatorio Astronomico, Vicolo Dell'Osservatorio 5, 
I-35122 Padova, Italy; arenzini@pd.astro.it}

\altaffiltext{7}{Code 667, NASA Goddard Space Flight Center, 
Greenbelt, MD 20771; allen.v.sweigart@nasa.gov, randy.a.kimble@nasa.gov}

\begin{abstract}

Using the Advanced Camera for Surveys on the {\it Hubble Space
Telescope}, we have obtained deep optical images reaching stars well
below the oldest main sequence turnoff in the spheroid, tidal stream,
and outer disk of the Andromeda Galaxy.  We have reconstructed the
star formation history in these fields by comparing their
color-magnitude diagrams to a grid of isochrones calibrated to
Galactic globular clusters observed in the same bands.  Each field
exhibits an extended star formation history, with many stars younger
than 10~Gyr but few younger than 4~Gyr.  Considered together, the star
counts, kinematics, and population characteristics of the spheroid
argue against some explanations for its intermediate-age, metal-rich
population, such as a significant contribution from stars residing in
the disk or a chance intersection with the stream's orbit.  Instead,
it is likely that this population is intrinsic to the inner spheroid,
whose highly-disturbed structure is clearly distinct from the
pressure-supported metal-poor halo that dominates farther from the
galaxy's center.  The stream and spheroid populations are similar, but
not identical, with the stream's mean age being $\sim$1~Gyr younger;
this similarity suggests that the inner spheroid is largely polluted
by material stripped from either the stream's progenitor or similar
objects.  The disk population is considerably younger and more
metal-rich than the stream and spheroid populations, but not as young
as the thin disk population of the solar neighborhood; instead, the
outer disk of Andromeda is dominated by stars of age 4--8~Gyr,
resembling the Milky Way's thick disk.  The disk data are inconsistent
with a population dominated by ages older than 10~Gyr, and in fact do
not require any stars older than 10~Gyr.

\end{abstract}

\keywords{galaxies: evolution -- galaxies: stellar content --
galaxies: halos -- galaxies: spiral -- galaxies: individual (M31)}

\section{Introduction}

One of the primary quests of observational astronomy is measuring the
formation history of structures ranging in scale from individual
galaxies to superclusters of galaxies.  However, a serious impediment
to this research is the fact that we live in a cosmological backwater.
The Local Group hosts only two giant spiral galaxies, the Milky Way
and Andromeda (M31, NGC~224), and no giant elliptical galaxies.  The
nearest galaxy groups to our own lie beyond 3 Mpc, with the closest
(the Maffei Group) being heavily reddened (Karachentsev 2005).

Given our rural setting, it is not surprising that our own Galaxy
drives the textbook picture of a giant spiral galaxy, with an ancient,
metal-poor halo (e.g., Ryan \& Norris 1991; VandenBerg 2000), an
ancient, metal-rich bulge (e.g., Zoccali et al.\ 2003 ; McWilliam \&
Rich 1994), and a disk hosting a wide range of ages and metallicities
(e.g., Fontaine et al.\ 2001; Ibukiyama \& Arimoto 2002).  However,
stellar population work in the Milky Way is often limited by
uncertainties in distance and reddening, and it is not even clear that
the Milky Way is representative of giant spiral galaxies in general.
Debate continues about the structure of the Milky Way system, how it
formed, and how its various substructures (halo, bulge, disk, globular
clusters, satellites, and tidal debris streams) formed with respect to
one another.  Physical processes possibly at work in forming the Milky
Way include rapid dissipative collapse in the early universe (Eggen et
al.\ 1962) and slower accretion of separate subclumps (Larson 1969;
Searle \& Zinn 1978).  More recent hierarchical models suggest that
spheroids form in a repetitive process during the mergers of galaxies
and protogalaxies, while disks form by slow accretion of gas between
merging events (e.g., White \& Frenk 1991).

Although hierarchical models based on cold dark matter (CDM) show
great success in reproducing the observable universe on scales larger
than 1 Mpc, these models predict many more dwarf galaxies than are
actually seen around the Milky Way (Moore et al.\ 1999).  This
discrepancy implies the existence of other mechanisms at work on small
scales.  For example, Bullock, Kravtsov, \& Weinberg (2000) suggested
that after the epoch of reionization, photoionization would suppress
gas accretion in small subhalos, keeping most of them dark-matter
dominated, and that a large fraction of those subhalos that did become
dwarf galaxies would be tidally disrupted into the halos of their
parent galaxies.  However, Grebel \& Gallagher (2004) argue that the
presence of ancient stars in all dwarf galaxies, along with their wide
variety of star formation histories, is evidence against a dominant
evolutionary effect from reionization.  Furthermore, Shetrone et al.\
(2003) demonstrated that chemical differences between nearby dSphs and
the Galactic halo imply that the halo is not comprised of populations like
those of present-day dSphs.  Whether or not accretion of dwarf
galaxies is the dominant source of stars in the halo, it is
likely that such galaxies do contribute, and at large galactocentric
distances their stars can remain in coherent orbital streams for 1 Gyr 
or more.  The discovery of the Sgr dwarf (Ibata et al.\ 1994) rekindled
interest in halo formation through accretion of dwarf galaxies,
leading to ambitious programs to map the spatial
distribution, kinematics, and chemical abundance in the halos of the
Milky Way (e.g., Morrison et al.\ 2000; Majewski et al.\ 2000) and
Andromeda (e.g., Ferguson et al.\ 2002; Guhathakurta et al.\ 2005).  A
spectacular example of this process has been found in Andromeda (Ibata
et al.\ 2001), which hosts a giant tidal stream extending several
degrees on the sky (McConnachie et al.\ 2003).  Indeed, the star count
map of Ferguson et al.\ (2002) shows complex substructure throughout
Andromeda, while also showing evidence for an underlying
smooth spheroid extending to large radii.

Besides the overabundant satellite problem, another issue with
hierarchical CDM models is their prediction that gas loses much of its
angular momentum during disk formation, resulting in theoretical disks
that are much smaller than those observed (Navarro \& Benz 1991).
Some have turned to warm dark matter (WDM) cosmologies to alleviate
this problem (e.g., Sommer-Larsen \& Dolgov 2001), but there is an
indication that angular momentum remains a problem in these models
(see Bullock, Kravtsov, \& Col$\acute{\rm i}$n 2002).  Alternatively,
the solution might lie in the inclusion of supernova feedback from
the earliest generation of stars; recent models that show promise in
this area predict that the bulk of the disk population was formed
relatively late, at $z \lesssim 1$ (Thacker \& Couchman 2001; Weil, Eke,
\& Efstathiou 1998), a prediction supported by panchromatic
surveys of large numbers of galaxies (Hammer et al.\ 2005).

As the nearest giant spiral galaxy to our own, Andromeda offers an
essential laboratory for studying the evolution of spiral galaxies.
Given our vantage point, one might even argue that it is a better
laboratory than our own Galaxy.  At 770~kpc (Freedman \& Madore 1990),
the stars in Andromeda all appear to be at approximately the same
distance, and at an inclination of $12^{\rm o}$ (de Vaucouleurs 1958),
its various structures can be studied somewhat independently.  We can
resolve Andromeda's old main sequence stars with the {\it Hubble Space
Telescope (HST)}, while the horizontal branch (HB) and upper red giant
branch (RGB) are accessible to observatories on the ground.  Recent
years have seen an enormous increase in observing time directed at
Andromeda, with deep pencil-beam surveys providing its star formation
history (e.g., Brown et al.\ 2003; Brown et al.\ 2006; Stephens et
al.\ 2003; Olsen et al.\ 2006) and shallow wide-field surveys
providing maps of its morphology, metallicity, and kinematics (e.g.,
Ibata et al.\ 2001; Ibata et al.\ 2004; Ibata et al.\ 2005;
McConnachie et al. 2003; Ferguson et al.\ 2002; Ferguson et al.\ 2005;
Kalirai et al.\ 2006b; Guhathakurta et al.\ 2005).

At first glance, Andromeda and the Milky Way appear to be very
similar; both are of similar Hubble type, luminosity, mass, and size
(van den Bergh 1992; van den Bergh 2000; Klypin, Zhao, \& Somerville
2002).  However, we have long known that the Andromeda spheroid is
very different from that of the Milky Way.  The first evidence came
from Mould \& Kristian (1986), who found that the mean metallicity in the
M31 halo, at 7 kpc on the minor axis, was surprisingly high
([m/H]$\approx -0.6$).  Pritchet \& van den Bergh (1994) subsequently
found that the halo surface brightness profile, out to a distance of
20 kpc on the minor axis, follows a de Vaucouleurs
exp$[-7.67(r/r_e)^{1/4}]$ profile instead of the $r^{-2}$ power law
expected for a canonical halo.  These results were extended by Reitzel
\& Guhathakurta (2002), Durrell et al.\ (2001, 2004), and Bellazzini
et al.\ (2003), who found that the high metallicity and de Vaucouleurs
profile continued out to distances of 20--30 kpc on the minor axis.
Recent surveys began probing M31 over much wider areas and much more
deeply.  Ferguson et al.\ (2002) mapped the density of bright RGB
stars over 25 square degrees of the galaxy, finding significant
substructure in the halo and outer disk.  With photometry extending
down to the oldest main sequence, Brown et al.\ (2003) reconstructed
the star formation history in the halo at 11 kpc on the minor axis,
and found a wide age distribution, with $\sim$30\% of the stars at
ages of 6--8 Gyr.  All of these studies suggested that the M31 halo is
dramatically different than the Milky Way halo, begging the question
``which is representative of large spiral galaxies?''  One possible
answer comes from Mouhcine et al.\ (2005), who found that the
metallicities of spiral halos correlate well with their parent galaxy
luminosities; the Milky Way halo falls well off this
metallicity-luminosity relation (being unusually metal-poor for the
parent galaxy mass), while the M31 halo appears representative for
large spiral galaxies.  It is unclear if this trend is due to a general
tendency for more massive galaxies to host a more dominant bulge,
ingest more satellites, and/or ingest larger satellites.

Recently, two independent groups (Guhathakurta et al.\ 2005; Irwin et
al.\ 2005) studying the outskirts of M31 found an extended stellar
halo that more closely resembles the halo of our own Milky Way.  This
extended halo begins to dominate beyond 30 kpc, where the minor-axis
surface-brightness profile transitions from a de Vaucouleurs law to an
$r^{-2.3}$ law.  From the colors in their photometric sample, Irwin et
al.\ (2005) concluded that the metallicity in the extended halo was as
high as it is in the region interior to 30 kpc, but Guhathakurta et
al.\ (2005), using a spectroscopically-confirmed sample extending
3 times farther out on the minor axis, found that this extended
halo is metal poor.  However, the existence of a metallicity gradient
was later confirmed by Kalirai et al.\ (2006b).  These discoveries can
lead to a confusion of terminology.  It seems straightforward to refer
to the inner few kpc as the bulge, and to the stars beyond 30 kpc as
the halo, but what about the stars at 5--30 kpc on the minor axis?
Before the discovery of the extended metal-poor halo, this population
of metal-rich stars was generally referred to as the halo, but it is
quite possible that this stellar population is more closely related to
the bulge.  Furthermore, there has been considerable debate about the
contribution of disk stars at these radii.  Kinematic studies indeed
show that M31 has an extended thick disk (Ibata et al.\ 2005).  The
minor-axis population at 11 kpc from the center has kinematics that
are inconsistent with a rotationally supported disk (Kalirai et al.\
2006b; Rich et al.\ in prep.), but the velocity dispersion is smaller
than might be expected for a purely pressure-supported stellar system.
Because the term ``spheroid'' normally refers to a structure that
includes the bulge and halo, we use the term here when referring to
the extraplanar stars at 5-30 kpc, merely to distinguish from those
regions that can be clearly labeled bulge (within $\sim$5 kpc of the
nucleus), disk (within $\sim$30~kpc on the major axis), and halo
(beyond $\sim$30~kpc on the minor axis).  However, our use of this
term is not intended to imply a smooth, relaxed, pressure-supported
structure.

To further understand the formation of Andromeda, we need to know the
star formation histories of its various structures.  To that end, we
have now obtained deep {\it HST} images in three fields, located in
the inner spheroid, outer disk, and the giant tidal stream.  All of
these images reach well below the oldest main sequence turnoff (MSTO)
in the galaxy, allowing a reconstruction of the entire star formation
history in each field.  Keck spectroscopy in each of these fields
provides additional kinematic information (Kalirai et al.\ 2006b;
Reitzel et al.\ 2006, in prep; Rich et al.\ 2006, in prep).  In
previous papers (Brown et al.\ 2003; Brown et al.\ 2006) we presented
the preliminary analysis of the spheroid and stream fields.  In this
paper, we present the detailed star formation histories in the
spheroid, stream, and outer disk of Andromeda.  In \S\ref{secobs}, we
describe our observing strategy and the data.  We describe the data
reduction in \S\ref{secred}, followed by the production of the
photometric catalogs in \S\ref{secphot}.  In \S\ref{secanal}, we
describe our analysis, which ranges from qualitative inspection of the
color-magnitude diagrams (CMDs) 
to quantitative fitting of the star formation histories,
including a full exploration of the possible systematic effects of our
assumptions.  In \S\ref{secdisc}, we discuss the implications of our
analysis.  The results of our study are summarized in \S\ref{secsumm}.

\section{Observations}
\label{secobs}

\begin{figure}[h]
\epsscale{1.2}
\plotone{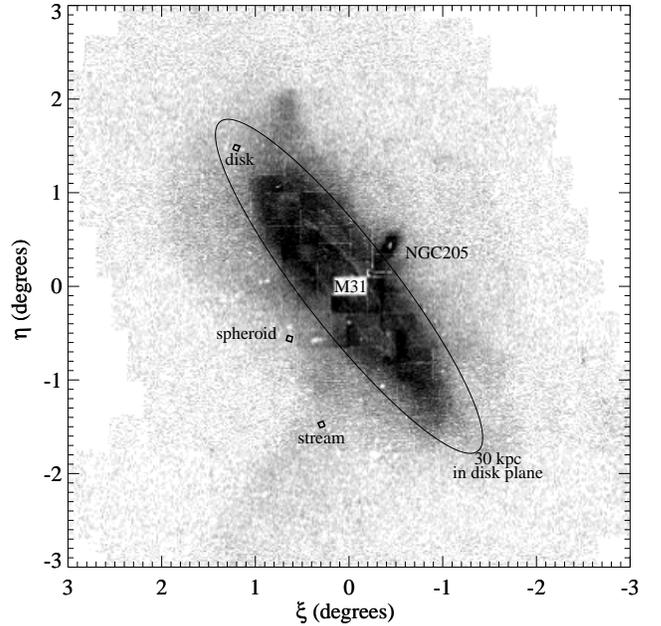}
\epsscale{1.0}
\vskip -0.1in
\caption{A map of stellar density in the Andromeda vicinity, 
from counts of RGB stars (Ferguson et al.\ 2002).  Appropriately scaled
and rotated boxes denote our three fields ({\it labeled}).  An ellipse
marks the area within 30 kpc of the galactic center in the inclined
disk plane ({\it labeled}).}
\label{mapfig}
\end{figure}

Using the Advanced Camera for Surveys (ACS; Ford et al.\ 1998) on {\it
HST}, we obtained deep optical images of three fields in M31: the
spheroid, outer disk, and tidal stream (Figure~\ref{mapfig};
Table~\ref{fieldtab}).  We used the F606W (broad $V$) and F814W ($I$)
filters on the Wide Field Camera (WFC).  The spheroid data, obtained
in the first {\it HST} observing cycle with ACS, reach their goal of
$\sim$1.5~mag below the oldest MSTO, while the stream and disk data,
obtained two years later, reach their goal of $\sim$1.0~mag below the
oldest MSTO.

\begin{table*}[t]
\begin{center}
\caption{Field Characteristics}
\begin{tabular}{lcccccc}
\tableline
      & R.A.    & Dec.    & log $N_{HI}$    & F606W  & F814W  &      \\
Field & (J2000) & (J2000) & (10$^{19}$ cm$^{-2}$) & (ksec)          & (ksec)         & Date \\
\tableline
spheroid & 00:46:07.1 & 40:42:39 & 19.3\tablenotemark{a}       & 139 & 161 & 2 Dec 2002 -- 1 Jan 2003 \\ 
stream   & 00:44:18.2 & 39:47:32 & $< 17.6$\tablenotemark{b} &  53 & 78 & 30 Aug 2004 -- 4 Oct 2004 \\
disk     & 00:49:08.6 & 42:45:02 & 20.6\tablenotemark{a}   &  53 & 78 & 11 Dec 2004 -- 18 Jan 2005 \\
\tableline
\multicolumn{7}{l}{$^{\rm a}$Braun et al.\ in prep; D. Thilker, private communication.} \\
\multicolumn{7}{l}{$^{\rm b}$Thilker et al.\ (2004).}\\
\label{fieldtab}
\end{tabular}
\end{center}
\end{table*}

The original spheroid program was proposed before the installation of
ACS on {\it HST}, and at that time, no images of M31 had reached
significantly below the level of the HB (3~mag brighter than an
old MSTO).  Given the uncertainties in this situation, our goal was an
ambitious depth that would unambiguously characterize the star
formation history in the spheroid.  With the proven capabilities of
ACS and the success of this program, we subsequently proposed a less
conservative approach as far as depth was concerned, giving up 0.5~mag
of depth for the ability to explore two fields without increasing
the size of the observing program.

Surface brightness was the primary driver in field selection.  There
are two competing factors in obtaining a CMD
appropriate for reconstructing the star formation history.  One wants
to maximize the number of stars in the CMD, to minimize contamination
from foreground stars and background galaxies, to minimize statistical
uncertainties in the characterization of the population, and to allow
the detection of subtle CMD features (e.g., small bursts due to
interaction of Andromeda with its satellites).  One also wants to
minimize the crowding, in order to maximize the accuracy of the
photometry for a given exposure time.  To explore these competing
factors, we created realistic simulations of ACS images under various
population assumptions, and found that the optimal crowding is
approximately one star of interest for every 50--100 resolution
elements; these results agree well with those of Renzini (1998).  The
native ACS/WFC pixel size is 0$^{\prime\prime}$.05, which is
approximately twice the width for critically sampling the point spread
function (PSF), so the number of resolution elements is roughly the
number of pixels.  This translates into $\sim$250,000 stars in an ACS
image, corresponding to a surface brightness $\mu_V \approx$26.3~mag
per square arcsec, which defines a roughly elliptical isophote around
M31 that provides the optimal crowding.

Fortuitously, the intersection of this isophote with the southern
minor axis falls near a globular cluster (SKHB-312) previously imaged
with the Wide Field Planetary Camera 2 (WFPC2) on {\it HST} (Holland
et al.\ 1997), so the exact position of our spheroid image was chosen
to place this cluster at the edge of our field.  Holland et al.\
(1996) determined the metallicity distribution in this field, and
showed that it was very similar to that observed in other fields
throughout the inner spheroid.  Although Holland et al.\ (1997)
reported a 10$^{\prime\prime}$ tidal radius for this cluster, we
placed it at the edge of our field in case deeper images revealed a
larger extent to the cluster that would contaminate a significant
fraction of the image and negatively impact the primary goal of
studying the field population.  Our photometry of SKHB-312 reached
well below its MSTO, revealing a cluster age of 10~Gyr (Brown et al.\
2004b).  Brown et al.\ (2004b) found no evidence for extended tidal
tails in the cluster, and so for the current study, we mask the area
within 15$^{\prime\prime}$ of the cluster center.  In the subsequent
observations of the tidal stream and outer disk, there was only a
candidate globular cluster (Bol D242; Galleti et al.\ 2004) near our
optimal position in the stream, and no known globular clusters near
our optimal position in the disk.  The exact location of the stream
field was chosen to include this candidate cluster (which subsequently
turned out to be a superposition of foreground stars), whereas the
exact location of the disk field was chosen to minimize the
contribution from the spheroid, based upon the disk/spheroid
decomposition of Walterbos \& Kennicutt (1988).  The surface brightness
in the stream field is $\approx$0.5~mag fainter than that in our
original spheroid program, while that in the disk field is
$\approx$0.1~mag brighter than that in our original spheroid program.
The hydrogen column density in the disk field is also much larger than
that in the spheroid and stream fields (see Table~\ref{fieldtab} for
$N_{HI}$ measured at each of our field positions).

Each exposure in a given bandpass was dithered so that no two
exposures placed a star on the same pixel.  Dithering smooths out
sensitivity variations across the detector, fills in the gap between
the two halves of the ACS/WFC detector, allows optimal sampling of the
PSF, and enables the removal of hot pixels.  Our dither pattern
employed three tiers of dithers to optimize the data quality.  The
first two tiers determined the nominal field position for one of our
visits in a given band (usually spanning two orbits), while the final
tier provided a 4-point dither pattern to optimally sample the PSF
within a visit.  The offsets in the first dither tier moved from -5 to
+10 pixels in X, with steps of 5 pixels, and from +60 to -120 pixels
in Y, with steps of 60 pixels, to place the detector gap at four
adjacent positions on the sky.  These first-tier offsets produce four
horizontal strips in our data where the field is underexposed by 25\%;
stars in these strips are ultimately discarded from our catalog, but
sampling the sky in the detector gap yields more accurate PSF-fitting
for the field because we have contiguous photometry for all of the
objects in the field.  The offsets in the second dither tier moved
-4.5, 0, or +4.5 pixels independently in X and Y, to smooth out
small-scale variations in detector response, plus a random fractional
pixel in X and Y, to avoid aliasing effects between the various
dithers, the pixel plate scale, and the geometric distortion.  The
offsets in the third tier were (0,0), (+1.5,0), (+1.5,+1.5), and
(0,+1.5) pixels in X and Y, to sample the PSF at twice the frequency
provided by a single exposure.

Each of these programs obtained brief exposures of Galactic star
clusters with the same filters on the ACS/WFC (Table~\ref{tabclus};
Brown et al.\ 2005).  The resulting CMDs provide empirical isochrones
that can be compared directly to the Andromeda CMDs and used to
calibrate the transformation of theoretical isochrones to the ACS
bandpasses.  These cluster observations, the empirical isochrones, and
the transformation of the theoretical Victoria-Regina Isochrones
(Vandenberg, Bergbusch, \& Dowler 2006) to the ACS bandpasses are
fully detailed by Brown et al.\ (2005).  We will use these empirical
isochrones and theoretical isochrones here, shifted to the Andromeda
reference frame by assuming a distance of 770~kpc (Freedman \& Madore
1990) and a reddening of $E(B-V)=0.08$~mag (Schlegel, Finkbeiner, \&
Davis 1998).  Over the region defined for fits to the star formation
history, the theoretical isochrones agree with the observed cluster
CMDs at the 0.02~mag level.

\begin{table}[b]
\begin{center}
\caption{Parameters$\tablenotemark{a}$ of Galactic clusters observed with ACS}
\begin{tabular}{llllr}
\tableline
                 &  $(m-M)_V$       & $E(B-V)$ &        & age \\
Name             &  (mag)           &  (mag)   & [Fe/H] & (Gyr)\\
\tableline
\tableline
NGC~6341 (M92)   & 14.60 & 0.023& $-2.14$ & 14.5\\
NGC~6752         & 13.17 & 0.055& $-1.54$ & 14.5\\
NGC~104 (47~Tuc) & 13.27 & 0.024& $-0.70$ & 12.5\\
NGC~5927         & 15.85 & 0.42 & $-0.37$ & 12.5\\
NGC~6528         & 16.31 & 0.55 & $+0.00$ & 12.5\\
NGC~6791         & 13.50 & 0.14 & $+0.30$ &  9.0\\
\tableline
\multicolumn{5}{l}{$^{\rm a}$Brown et al.\ 2005 and references therein.}\\
\end{tabular}
\label{tabclus}
\end{center}
\end{table}

\begin{figure*}[t]
\epsscale{1.1}
\plotone{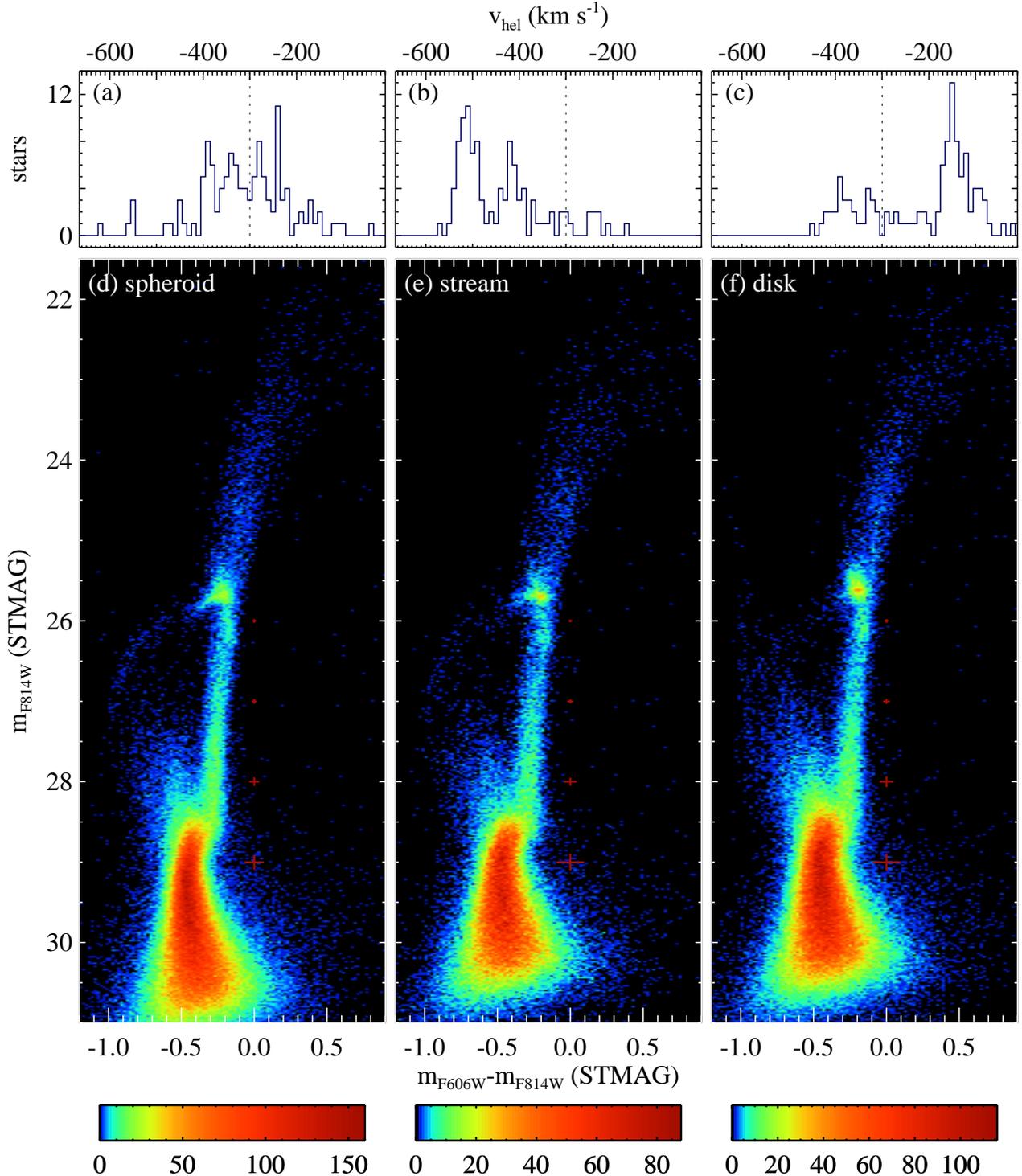}
\epsscale{1.0}
\caption{The velocities ({\it upper panels}) and CMDs ({\it lower
panels}; Hess diagrams at a logarithmic stretch) for our three fields, with
representative error bars ({\it red crosses} to the right of the
stellar locus).  Scales under each CMD indicate the number of stars
per bin.  The velocity histogram for our spheroid field {\it (a)}
shows a broad distribution ($\sigma \approx 80$ km s$^{-1}$) at the
Andromeda systemic velocity ({\it dashed line}), with no significant
contribution from a disk or single stream (Rich et al.\ in prep.).
The stream field ({\it b}) is dominated by two kinematically-cold
($\sigma \approx 15$ km s$^{-1}$) stream components, falling toward
the observer from behind the galaxy, with the remainder in a broader
spheroid component at the systemic velocity (Kalirai et al.\ 2006b).
The disk field ({\it c}) is dominated by a narrow disk component,
redshifted due to the rotation of the disk, with the remainder in a
broader spheroid component at the systemic velocity (Reitzel et al.\
in prep).  The populations in the spheroid ({\it d}) and stream ({\it
e}) look remarkably similar (Brown et al.\ 2006), although the
spheroid data are 0.5 mag deeper.  The disk, in contrast, shows a
younger population (note the red clump HB morphology and the blue
plume above the dominant MSTO) at higher mean metallicity (note the
redder RGB).  Figures~\ref{compare} and \ref{subtract} highlight the
distinctions among the three populations.}
\label{cmdsvels}
\end{figure*}

A sample of bright RGB stars in our three fields has been observed
spectroscopically with Keck, providing crucial kinematic context for
each field.  The velocity data in all three fields are presented by
Kalirai et al.\ (2006b), but the focus of that paper is the kinematic
structure of the tidal stream.  Rich et al.\ (2006, in prep.)  will
focus on the kinematic structure of the spheroid, while Reitzel et
al.\ (2006, in prep.) will focus on the kinematic structure in the
outer disk.  The velocity information in each of our fields is
presented in Figure~\ref{cmdsvels}.  The velocities in the spheroid
field show a broad distribution, with no dominant contribution from a disk
or a single stream.  The velocities in the stream field show it to be
dominated ($\approx 3/4$) by stars moving in two narrow stream
components, with the remainder in the spheroid.  The velocities in the
disk field show it to be dominated ($\approx 2/3$) by stars moving in
a disk component, with the remainder in the spheroid.

\section{Data Reduction}
\label{secred}

If calibrated data are retrieved from the {\it HST} archive as soon as
they are available, they will generally not have the best dark and
bias subtractions, because those calibration products are created
weeks later from a contemporaneous set of data that was obtained in
the days surrounding a given observation.  Thus, months after these
observations, we re-retrieved the images, yielding data with the
latest ACS pipeline calibration, including an appropriate dark
subtraction, flat-field, and bias correction; these are the ``FLT''
files in the ACS pipeline.  We then subtracted an iteratively
sigma-clipped median sky level from each quadrant of each image to
avoid an unnecessary increase in image noise during the later
coaddition of the images; this also corrects for small
quadrant-dependent bias residuals.  We used the PyRAF DRIZZLE package
(Fruchter \& Hook 2002) to register the individual images, correct for
geometric distortion and plate scale variations, reject cosmic rays,
and coadd all of the frames in a given bandpass.  The geometric
distortion correction employed the coefficients provided by the ACS
calibration pipeline for each image, which include both the general
geometric distortion and also the time-varying plate scale changes due
to velocity aberration.  As part of the drizzle process, the images
were resampled to a plate scale of 0$^{\prime\prime}$.03 pixel$^{-1}$.
Residual shifts, rotations, and plate scale variations were corrected
as part of our registration process, and thus our reduction is immune
to the software errors that have sometimes caused registration and
photometry problems in the MULTIDRIZZLE software (although note that
we do not use MULTIDRIZZLE in our reduction).

To register the images, we drizzled the images to individual output
frames that were corrected for geometric distortion and velocity
aberration, with relative shifts determined by the pointing
information in the image headers (closely matching our commanded
dither pattern).  The positions of $\sim$10,000 relatively bright
stars, well-detected in the individual images, were then measured in
each image through the entire image stack, using an iterative fit of a
Gaussian profile to each star.  These stellar positions were used to
refine the offsets (deviations from the guide star offsets), rotations
(deviations from the fixed orientation requested), and plate scale
changes (telescope breathing and residual velocity aberration).  Using
the refined knowledge of the relative astrometry, we re-drizzled the
images to individual output frames.  These refinements to the offsets,
rotations, and scales are iterated until the positions of the bright
stars in the individual images are aligned to better than 0.01 pixels.

\begin{figure*}[t]
\epsscale{1.1}
\plotone{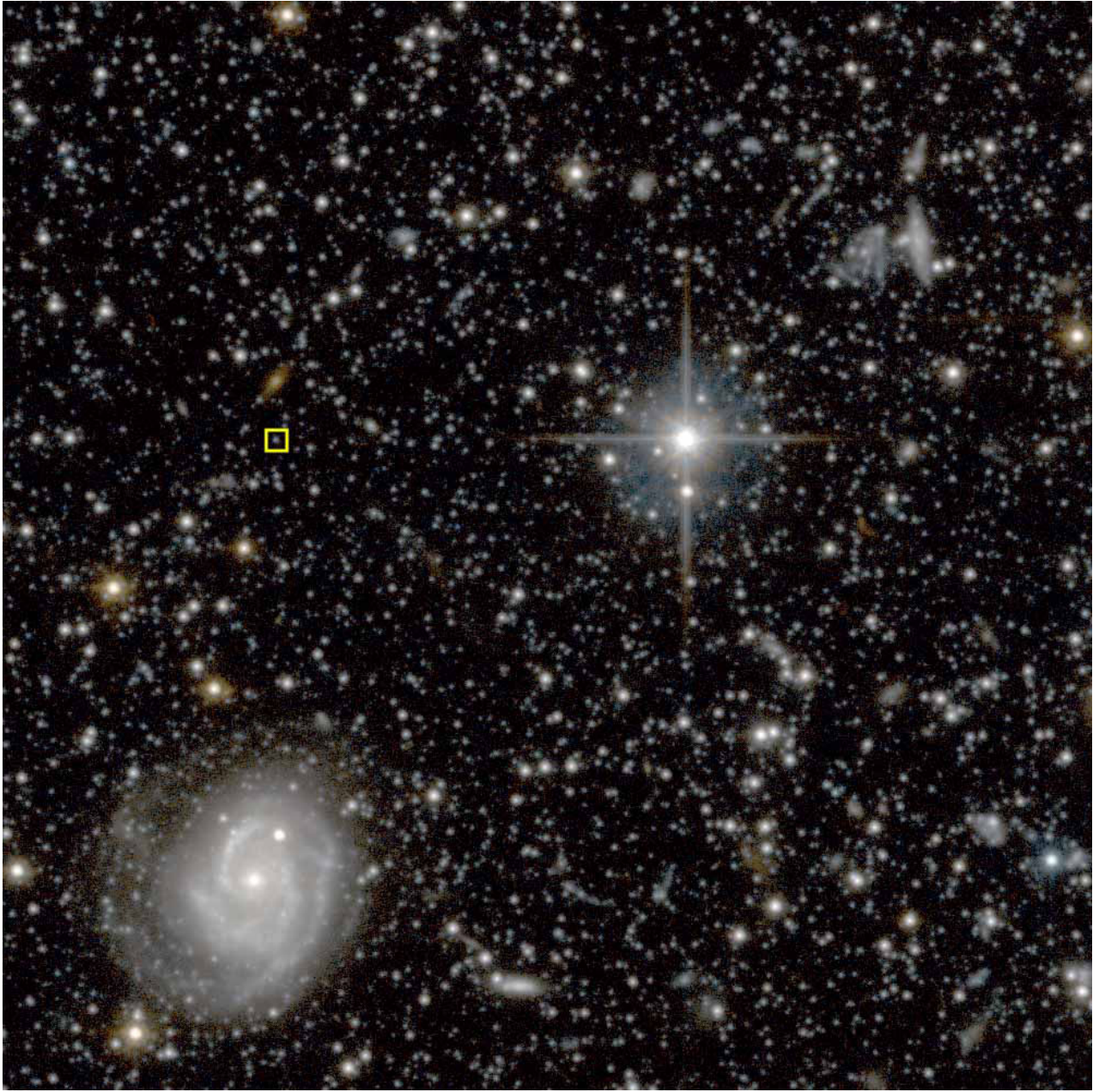}
\epsscale{1.0}
\caption{A $30^{\prime\prime}\times30^{\prime\prime}$ subsection of
our spheroid field, with a logarithmic stretch.  The blue channel
comes from the F606W exposure, the red channel comes from the F814W
exposure, and the green channel comes from the sum of those two bands.
A yellow box marks a star near the MSTO, at $m_{F814}=29$~mag,
$m_{F606W}-m_{F814W} = -0.45$~mag; it is clear that stars are detected
well below this point.  This subsection was chosen to give examples of
the types of objects in our field -- Galactic foreground stars (with
diffraction spikes), background galaxies, and Andromeda stars -- but
it is not a typical patch.  Most of the image is dominated by
Andromeda stars.}
\label{image}
\end{figure*}

Next we created masks of cosmic rays and problematic pixels.  Although
saturated pixels are masked in the data quality array (along with some
of the hot and dead pixels), saturation is only an issue for the
handful of bright foreground stars and not stars in M31; in our
half-orbit exposures, a star would have to be brighter than 20.6~mag
in either bandpass and well-centered in a pixel for saturation to
occur.  To create our masks, we first calculated a clipped median of
all the images in a given band, resulting in our first pass at the
deep image in that band.  The first-pass images were then
reverse-drizzled (or ``blotted'' in the drizzle nomenclature) back to
the original frame of the individual images.  The comparison of these
blotted images with each FLT image enabled the creation of masks for
the cosmic rays, self-annealed pixels (oversubtracted by the dark
calibration), and short-term transient warm and hot pixels not
corrected by the contemporaneous dark calibration.  To create a
complete mask for each frame, these custom masks were combined with
the pipeline-provided data quality masks in which we include all
pixels flagged for any reason.  The masked images were then coadded,
with weighting by exposure time, to create a second-pass deep image in
each band.  Because this iteration was significantly improved over the
first pass (median image), these second-pass deep images were then
blotted back to the original frames of the individual images, to
enable refinement of the image masks.  The frames were coadded a third
time to create the final image in each bandpass.  We then added a flat
sky component to each final image, representing the exposure-weighted
mean of the sky background subtracted from the individual frames, to
ensure that the counting statistics were appropriate in the
subsequent photometric reduction.  In Figure~\ref{image}, we show a
false-color $30^{\prime\prime}\times30^{\prime\prime}$ subsection of
our spheroid field, combining the images in F606W and F814W filters.

Although this process was repeated on the full set of data for each
field, we also applied the process to a subset of the spheroid data
chosen to match the shorter exposure times in the stream and disk.  In
this paper, the fits to the star formation history in the spheroid
utilize the full dataset, but the shallower version of the spheroid
data is useful for making a fair comparison of the CMDs of the three
fields.  Spheroid CMDs utilizing this subset of the data are labeled
``matched.''

\section{Photometry}
\label{secphot}

We used the DAOPHOT-II (Stetson 1987) PSF-fitting package to obtain
photometry of each field.  Empirical PSFs were created from the images
using the most isolated and well-exposed stars, with a radius of 23
pixels (0$^{\prime\prime}$.69).  We first performed an initial pass of
object detection and aperture photometry.  The resulting object
catalog was then clipped to retain only those well-detected stars that
fell within the dominant stellar locus in the CMD (rejecting outliers
that are obvious blends and background galaxies), while also avoiding
those stars near the tip of the RGB and in the instability strip
(which can be variable and thus have PSFs adversely affected by the
cosmic-ray cleaning).  We then screened for stars with relatively
bright neighbors; to be a valid PSF star, all neighbors within 15
pixels must be at least 3~mag fainter than the PSF star, and all
neighbors within 22 pixels must be at least 2~mag fainter than the PSF
star.  Finally, we created a SExtractor (Bertin \& Arnouts 1996) map
of the thousands of background galaxies in our image, and removed from
our list of PSF candidates any star within 23 pixels of an extended
object.  We also removed stars within 23 pixels of an image border or
a bad pixel (e.g., due to charge bleeding from saturated stars).  This
resulted in $\sim$2000 PSF candidates per field in the disk and
spheroid and $\sim$1600 PSF candidates in the stream field.  These PSF
stars were then passed through an iterative process to create the
empirical PSF for each bandpass in each field.  In the first pass, we
used the PSF of Brown et al.\ (2003) as an initial guess at the
current PSF.  We fitted this PSF to the catalog of stars, subtracted
those stars, and then performed a new round of object detection on the
residual image to find the fainter stars uncovered.  We added these
stars to the catalog, and repeated the PSF-fitting photometry for the
entire catalog.  Stars neighboring each PSF star were then subtracted,
and a new PSF was then created.  During this process, we eliminated PSF
stars whose fits were high $\chi^2$ outliers (as reported by
DAOPHOT-II).  Also PSF stars were cut from the list if PSF subtraction
uncovered close stellar blends with the PSF star or revealed
underlying deviant pixel artifacts.  Using such a large number of PSF
stars, we are able to compare the morphology of the PSF-subtracted
residuals to those of nearby PSF stars and reject any with morphology
deviating from the pattern in that particular region of the image.
This process was iterated, each time increasing the allowed degree of
spatial variability in the fit, starting from a spatially-constant PSF
and ending at a third-order polynomial variation with field position.
This degree of variability is an advantage of the stand-alone
DAOPHOT-II code, as the IRAF version only allows a second-order trace
of the strong spatial variation in the ACS/WFC PSF.  Once we stopped
increasing the degree of spatial variability in the PSF, we iterated two
more times, purging problematic PSF stars after each new round of PSF
production, until an accurate spatially-varying PSF was created for
each bandpass in each field.  In the end, $\sim$1600 stars were used
to create the PSF in the disk and spheroid fields, while $\sim$1400
stars were used in the stream field.

With the empirical PSFs in hand, we performed PSF-fitting photometry
on the images to create the catalog of stars in each field.  First, we
used the ``find'' routine in the DAOPHOT-II package and a 5$\sigma$
detection threshold on the sum of the F606W and F814W images, to
create the initial pass at object detection.  After making an initial
estimate of magnitudes for the catalog with a round of aperture
photometry in each band, the catalog was cleaned of PSF substructure
misidentified as stars in the vicinity of well-exposed stars.  The
aperture photometry was then used as the starting point for
PSF-fitting photometry.  After these stars were fitted and subtracted
from each image, we summed the residual images in F606W and F814W to
create a new detection image, which was again fed to the find routine
with a 5$\sigma$ threshold.  The detections in this second pass were
often noise residuals from the subtraction of the stars in the first
pass, or other image artifacts revealed by the subtraction.  Our first
screen of these artifacts was made using the sharpness measurement
produced by the find algorithm.  We cut 4$\sigma$ outliers in the
sharpness distribution.  We then screened artifacts from bright star
residuals (consistent with Poisson noise) by removing detections
within a given magnitude-dependent radius of stars in the first-pass
catalog; the radius is magnitude-dependent because for fainter stars,
the residuals approach sky noise at a smaller radial distance from the
center of the PSF. We obtained aperture and then PSF-fitting
photometry of the second-pass stars, without re-centering, using the
original F606W and F814W images (i.e., the deep drizzled images as
they stood before any PSF subtractions).  The first-pass and
second-pass star lists were then combined, and another run of
PSF-fitting photometry (now allowing recentering) was performed on the
original F606W and F814W images.  DAOPHOT-II reports a goodness-of-fit
($\chi^2$) statistic for the PSF fits to each star; we analyzed the
distribution of this $\chi^2$ statistic as a function of stellar
magnitude, and marked those objects in the deep images that had high
$\chi^2$ values.  Inspection of the marked images showed that these
outliers were primarily due to close blends, PSF artifacts (e.g.,
diffraction spikes), and/or objects superimposed on background
galaxies (which include both true Andromeda stars superimposed on
background galaxies and substructure within background galaxies
incorrectly identified as stars).  We clipped from the catalog these
outliers in the $\chi^2$ distribution.  This clipping is responsible
for much of the improvement between the CMDs shown here and that
shown in the preliminary publication of the spheroid CMD (Brown et
al. 2003).  Finally, we discarded those stars falling in parts of the
image without the full exposure (due to dithering the image edges and
the detector gap).  Note that our artificial star tests (discussed
below) included all of the same processes and evaluations used in the
process that created the photometric catalog, so that any rejection of
real stars is reproduced in the simulated CMDs.

The PSF-fitting photometry was put on an absolute magnitude scale by
normalizing to aperture photometry on the brightest stars.  That
aperture photometry was itself put on an absolute magnitude scale
using aperture corrections determined from TinyTim (Krist 1995) models
of the ACS PSF.  The aperture corrections were verified with
observations of the standard star EGGR 102 (a $V=12.8$~mag DA white
dwarf) in the same filters; the agreement between the standard star
photometry and the TinyTim model is at the 1\% level.  In
Figure~\ref{cmdsvels}, we show the CMD for each field at its full
depth, along with the associated velocity distribution of RGB stars in
each field (Kalirai et al.\ 2006b; Rich et al.\ in prep; Reitzel et
al.\ in prep.).  Due to the large numbers of stars in each field, a
traditional CMD (with a point for every star) is saturated and
difficult to interpret; instead, we have binned the data into Hess
diagrams, with shading indicating the number of stars per bin.  The
same logarithmic stretch (characterized by the scales under each CMD)
spans the full range of stellar density in each CMD, but that range
varies from field to field given the variation in surface brightness
and observing depth.  The stretch was chosen to reveal both the subtle
and gross properties of each population.  We also plot representative
errors bars in each CMD, measured by taking the standard deviation
between the input and output values for the given color and magnitude
in our artificial star tests (discussed below); note that crowding
is the dominant source of scatter in each bandpass, which causes
photometric errors to be larger in either $m_{F606W}$ or $m_{F814W}$
than in $m_{F606W}-m_{F814W}$.

Our photometry is in the STMAG system: $m = -2.5 \times $~log$_{10}
f_\lambda -21.1$~mag, where $f_\lambda = $ e$^- \times {\rm
PHOTFLAM/EXPTIME}$, EXPTIME is the exposure time, and PHOTFLAM is
$7.906 \times 10^{-20}$ erg s$^{-1}$ cm$^{-2}$ \AA$^{-1}$ / (e$^{-}$
s$^{-1}$) for the F606W filter and $7.072 \times 10^{-20}$ erg
qs$^{-1}$ cm$^{-2}$ \AA$^{-1}$ / (e$^{-}$ s$^{-1}$) for the F814W
filter.  The STMAG system is a convenient system because it is
referenced to an unambiguous flat $f_\lambda$ spectrum; an object with
$f_\lambda = 3.63 \times 10^{-9}$ erg s$^{-1}$ cm$^{-2}$ \AA$^{-1}$
has a magnitude of 0 in every filter.  Another convenient and
unambiguous system that is widely used is the ABMAG system: $m=-2.5
\times $~log$_{10} f_\nu -48.6$~mag; it is referenced to a flat
$f_\nu$ spectrum, such that an object with $f_\nu = 3.63 \times
10^{-20}$ erg s$^{-1}$ cm$^{-2}$ Hz$^{-1}$ has a magnitude of 0 in
every filter.  It is thus trivial and unambiguous to convert any of
the data presented herein from STMAG to ABMAG: for F606W,
ABMAG~=~STMAG~$-$~0.169~mag, and for F814W, ABMAG~=~STMAG~$-$~0.840~mag.
Although our photometry could be transformed to ground magnitude
systems (e.g., Johnson $V$ and Cousins $I$) for comparison to
theoretical isochrones as well as other data in the literature, such
transformations always introduce significant systematic errors (see
Sirianni et al.\ 2005).  Instead of converting {\it HST} data to
ground bandpasses, it is preferable to produce models in one of the
{\it HST} instrument magnitude systems, in either STMAG or ABMAG.

Brown et al.\ (2006) found excellent agreement between the HB and RGB
distributions in the stream and spheroid populations if the stream is
assumed to be 0.03~mag (11 kpc) more distant than the spheroid. The
sense of the offset in luminosity is in agreement with the velocities
of the stars in the stream, which imply that the stream is falling into
Andromeda from behind it (Figure~\ref{cmdsvels}b; see also McConnachie
et al.\ 2003).  Brown et al.\ (2006) also found a 0.014~mag offset in
color between the stream and spheroid data, which is well within the
uncertainties in calibration and reddening.  Thus, we shifted the stream
CMD 0.03~mag brighter and 0.014~mag to the red, to put it in the same
frame of reference as the spheroid data.  These shifts are very small,
and make very little difference to the CMDs displayed herein or to the
various fits to the stream data, but we apply these shifts because
they are appropriate to the best of our knowledge.  The distinctions
between the spheroid and disk CMDs are far larger than the calibration
and reddening uncertainties, and in fact no single shift in color and
luminosity can align the features of the disk and spheroid CMDs.
Thus, to the best of our knowledge, the distinctions between the disk
and spheroid data are physical, and so the disk data are analyzed
without modification.

It is worth noting the implications of our shifts to the stream data
if these shifts are entirely due to a difference in extinction.  The
Schlegel et al.\ (1998) extinction map gives $E(B-V)=0.08$~mag at our
spheroid and disk positions and $E(B-V) =0.05$~mag at our stream
position, but this variation is within the uncertainties for their
map, which are generally $\sim 0.02$~mag in random fields and a bit
higher near Local Group galaxies.  At 3,500~$\le T_{\rm eff}
\le$~35,000~K, synthetic spectra folded through the ACS and ground
bandpasses imply $E(m_{F606W}-m_{F814W}) \approx E(B-V)$.  So, if we
took the map at face value, we would shift the stream data 0.03~mag to
the red and 0.05~mag fainter, to put the stream data in the same
extinction reference frame as the spheroid data.  However, a 0.03~mag
shift to the red is larger than the 0.014~mag required to align the
stream and spheroid color distributions at the HB and RGB.  Given the
uncertainties in the extinction map, we could instead shift the stream
data 0.014~mag to the red and 0.02~mag fainter.  This would align the
color distributions of the stream and spheroid at the HB and RGB, but
the stream HB would be $0.03 + 0.02 = 0.05$~mag fainter than the
spheroid HB, implying that the stream distance modulus in our field is
0.05~mag larger than the spheroid distance modulus.  In any case,
given that the calibration uncertainties for the $m_{F606}-m_{F814W}$
color are also at the same level as the color shift, it is not
appropriate to read too deeply into these small shifts in color and
magnitude between the fields.

Damage to the CCDs due to radiation in space leads to charge transfer
inefficiency (CTI), a problem that is particularly noticeable in
large-format CCDs.  CTI causes stars to appear fainter than they
actually are.  The ACS WFC detector consists of two chips, with $4096
\times 2048$ imaging pixels each, separated by a small gap.  Each CCD
is read out through two serial amplifiers, with 24 physical pixels of
leading serial overscan for each and 20 rows of trailing virtual
overscan in the parallel clocking direction, yielding a final
downlinked image format of $4144 \times 2068$ for each CCD.  Stars
that fall closer to the gap undergo more parallel transfers when the
detector is read, and thus suffer from more charge loss due to CTI
(for these CCDs, at the ACS operating temperature and clocking rates,
the CTI effects after radiation exposure are much more significant in
the parallel clocking direction than in the serial).  The CTI
correction is approximately linear with the position of a star
relative to the gap, and approximately linear with the age of the
detector.  The correction is larger for faint stars and smaller when
there is a significant background.  Our spheroid field was observed
shortly after the ACS launch, while the stream and disk fields were
observed two years later.  The standard CTI correction (Riess \& Mack
2005) was derived for brighter stars with lower backgrounds than the
situation in our images.  Thus, it is somewhat uncertain whether or
not one should extrapolate these CTI corrections into the regime of
our data, which includes the deepest stellar photometry obtained with
{\it HST} to date.  Fortunately, CTI does not appear to be a significant
problem in our images.  We checked the effects of CTI by constructing
CMDs of stars extracted from a range of horizontal
bands across the image.  The CMD includes two horizontal features
separated by approximately 3~mag in luminosity: the HB and
the subgiant branch (SGB).  The luminosity of each of these features
can be determined by taking a vertical cut through the CMD in the
vicinity of each feature (using a region restricted in color to avoid other
evolutionary phases). If CTI were a significant problem in our
data, one would expect the luminosity offset between the HB and SGB to
vary by a few hundredths of a magnitude as a function of vertical
position in the image, given the intensity of the sources and the
observed sky background.  In reality, we find that this offset varies by
$\lesssim$0.001~mag across the image.  Thus, there are probably
additional factors contributing to the CTI mitigation, besides the sky
background of $\sim$100 counts per pixel.  Because the images are
crowded with stars and background galaxies, most stars are clocked
across pixels where the charge traps have already been filled by other
sources.  Given the lack of evidence for CTI, we do not attempt a CTI
correction.  Note that any CTI correction applied to these data would
tend to make the stellar populations look slightly younger, because
the fainter main sequence and subgiant stars would have a larger
correction than the brighter HB stars.

We next performed extensive artificial star tests to characterize the
completeness and photometric scatter as a function of color and
magnitude in each field.  These tests required months of computations
on a dedicated cluster of 10 processors.  In all, 5 million artificial
stars were added to each field and blindly recovered, with these stars
spanning the full range of color and magnitude populated by the
stellar locus.  These stars were added in 1000 passes with 5000 stars
per pass, to avoid significantly increasing the crowding in the
images.  The artificial stars were blindly recovered with a process
identical to that used for the photometric catalog.  The completeness
exceeds 80\% at $m_{F814W} \leq 30.5$~mag in the spheroid data, and
exceeds 80\% at $m_{F814W} \leq 30.0$~mag in the disk and stream data,
but it drops off rapidly below these magnitudes.  These limits drive the
faint limit of the region we fit for the star formation history.  Note
that the images detect stars significantly fainter than those
presented in the CMDs presented here; compared to the reduction of
Brown et al.\ (2003), the catalog depth and completeness have been
somewhat reduced by the higher detection threshold and rigorous
cleaning process we have employed here.

\section{Analysis}
\label{secanal}

\subsection{Inspection of the Color-Magnitude Diagrams}
\label{secinspec}

Before turning to the quantitative fitting of the CMDs, much can be
learned from simple visual inspection.  The CMD for the population in
each field is shown in Figure~\ref{cmdsvels}.  At first glance, all
three CMDs look remarkably similar, even though the populations have
distinct kinematics.  All of them show a broad RGB, indicative of a
wide metallicity range.  In each field, the majority of the stars
between the MSTO and the base of the RGB are clustered in a tight
locus.  Given the spread in metallicity, this tight SGB locus
indicates a wide range in age, with younger stars generally more
metal-rich than older stars.  A minority population of stars appears
in a blue plume above the MSTO, representing a young population with a
wide range of metallicities.  We return to the SGB and blue plume
below.  Each of the fields has a well-defined HB, although the HB in
the disk field is largely restricted to a red clump, while in the
stream and spheroid $\sim$10\% of the HB stars fall on the blue end of
the HB.  None of the fields have an extended hot HB, as seen in massive
Galactic globular clusters spanning a wide range in metallicity (e.g.,
M19, at [Fe/H]~=~$-1.68$, and NGC6441, at [Fe/H]~=~$-0.53$; Piotto et
al.\ 1999; Rich et al.\ 1997).  Instead, the blue HB, when present,
very closely resembles that of typical metal-poor clusters, such as
M92, at [Fe/H]~=~$-2.1$ (see Brown et al.\ 2003).  The RGB luminosity
function bump is prominent in each CMD, at a luminosity
$\sim$0.5~mag fainter than the red end of the HB; this bump is a
metallicity indicator, becoming fainter (relative to the HB) at higher
metallicities, and in all three fields its spread in luminosity is
another indication of a spread in metallicity.  None of the CMDs shows
multiple discrete turnoffs, as might be expected from pulses of star
formation.

\begin{figure*}[t]
\epsscale{1.1}
\plotone{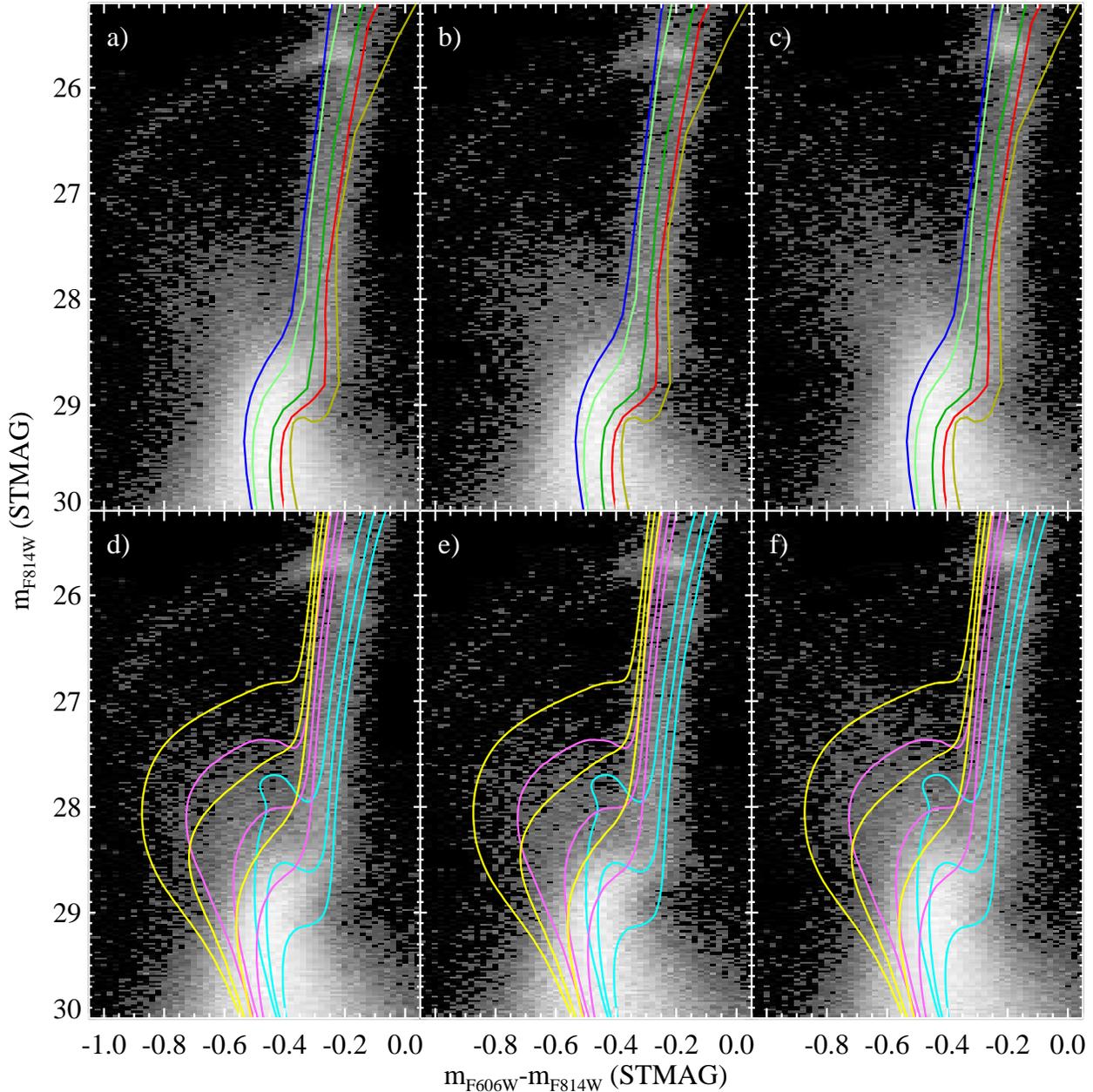}
\epsscale{1.0}
\caption{{\it Top panels:} The CMDs of the spheroid ($a$), stream
($b$) and disk ($c$), compared to the ridge lines of the Galactic globular
clusters in Table~\ref{tabclus} ({\it colored curves}).  
The Andromeda data are shown as Hess diagrams with the same binning
used in Figure~\ref{cmdsvels}, but over a narrower range of 
color and luminosity.  The ridge
lines shift redward with increasing cluster metallicity.  {\it Bottom
panels:} The CMDs of the spheroid ($d$), stream ($e$) and disk ($f$),
compared to isochrones at [Fe/H]~=~$-2$ ({\it yellow curves}), $-1$
({\it pink curves}), and 0 ({\it light blue curves}), and ages of 3,
8, and 13~Gyr (running from left to right for each color).  It is
clear that the old ($> 10$~Gyr) populations in these fields must be
predominantly metal-poor ([Fe/H]$\leq -1$), and that most of the
metal-rich populations ([Fe/H]$> -1$) must be of intermediate age
($\sim 6$--8~Gyr).}
\label{comparegrid}
\end{figure*}

In Figure~\ref{comparegrid}, we compare the CMDs to our globular
cluster fiducials (Table~\ref{tabclus}; Brown et al.\ 2005).  Due to
their wide range of metallicities, the clusters span most of the
RGB width in the M31 CMDs.  However, because the clusters are old,
there is an obvious trend for the MSTO and SGB in the more metal-rich
clusters to be too faint relative to those features in the M31 CMDs.
In the bottom panels of Figure~\ref{comparegrid}, we show a comparison
of the M31 CMDs to calibrated isochrones at three different ages (3,
8, and 13 Gyr) and three different metallicities ([Fe/H]~=~0, $-1$,
and $-2$).  It is clear that the old ($> 10$~Gyr) populations in these
fields must be predominantly metal-poor ([Fe/H]$\leq -1$), and that
the metal-rich populations ([Fe/H]$> -1$) must be of intermediate age
($\sim 6$--8~Gyr).  An old metal-rich population would have a MSTO
much redder and fainter than observed, while an intermediate-age
metal-poor population would have a MSTO much bluer and brighter than
observed.  That said, there is a minority population of young stars
spanning a wide range in metallicity, with the brightest and bluest
stars in the plume matched by the 3~Gyr isochrone at [Fe/H]~=~$-2$.

\begin{figure*}[t]
\epsscale{1.1}
\plotone{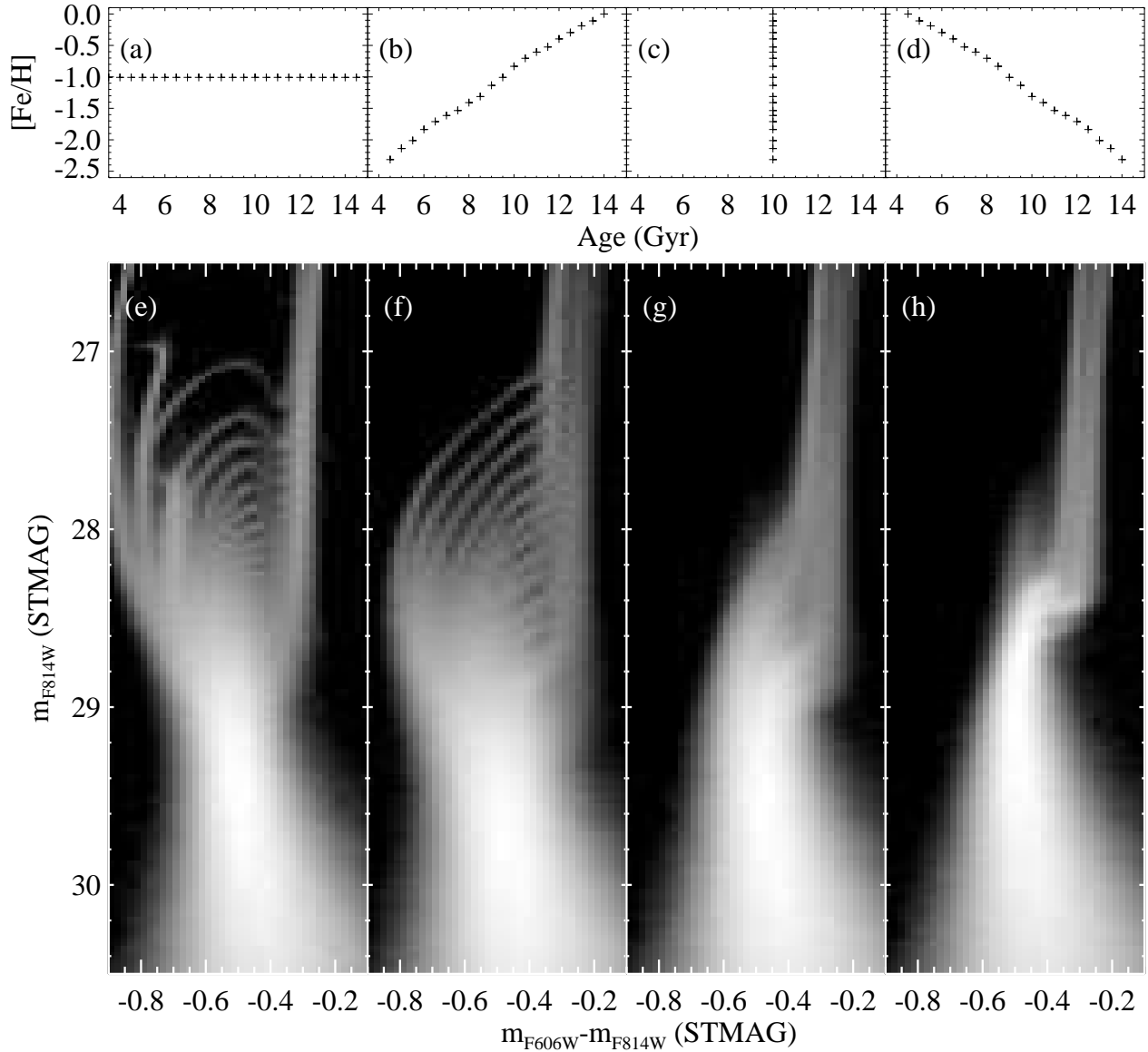}
\epsscale{1.0}
\caption{{\it Top panels:} Four hypothetical populations of stars.  
In each population, the stars are equally distributed among 20 isochrones
with distinct distributions in age and metallicity.
{\it Bottom panels:} Model CMDs for these hypothetical
populations, with the observational errors and completeness of our
spheroid data, shown at a logarithmic stretch.}
\label{sgbwidth}
\end{figure*}

The implications of the SGB distribution warrant additional
discussion.  The isochrones in Figure~\ref{comparegrid} show that the
luminosity of the SGB decreases with either increasing age or
increasing metallicity.  Thus, different age-metallicity relations for
the stars in our CMDs would be expected to produce different
luminosity distributions across the SGB.  To evaluate the implications
of this constraint, we show in Figure~\ref{sgbwidth} hypothetical
populations of stars in the vicinity of the SGB as they would appear
if observed under the same conditions as in our spheroid field.  The
upper panels present the age-metallicity relations of the isochrones
employed to construct each model population, with the stars divided
equally among the isochrones.  The lower panels show the corresponding
CMDs resulting from these hypothetical populations.  Even with a very
wide range in age, a single metallicity does not reproduce the width
of the RGB (panels $a$ and $e$); this is because the RGB is far more
sensitive to metallicity than to age.  Moreover, the SGB luminosity
distribution is much wider than observed.  If one has old metal-rich
stars and young metal-poor stars (panels $b$ and $f$), the RGB becomes
much wider, but the SGB luminosity distribution is still much wider
than observed.  If all of the stars are at a single age (panels $c$
and $g$), the SGB narrows, but it is still wider than the SGB observed
in our fields.  It is only when one has young metal-rich stars and old
metal-poor stars (panels $d$ and $h$) that the SGB locus becomes very
tight and horizontal, as observed for the dominant populations in our
three CMDs, while at the same time reproducing a wide RGB.  Because
the RGB is more sensitive to metallicity than to age, while the MSTO
is very sensitive to both, one is able to break the age-metallicity
degeneracy in studies employing this region of the CMD.  Note
that relatively young and metal-poor stars (panels $a$, $b$,
$e$, and $f$) are needed to explain the brightest and bluest stars in
the blue plume of our observed CMDs.

The similarities at the HB and RGB between the stream and spheroid
imply that these populations have very similar metallicity distributions,
at least at the positions of our fields (Brown et al.\ 2006).  Much
farther out in the galaxy (31 kpc from the center), Guhathakurta et
al.\ (2006) found that the stream was more metal-rich than the
surrounding spheroid, but this finding is not inconsistent with our
results.  Kalirai et al.\ (2006a) have shown that the spheroid of
Andromeda has a metallicity gradient, such that it is significantly
more metal poor at 30 kpc than it is close to the galaxy's center.
Our finding of similar metallicities between the stream and
spheroid in our interior fields, when combined with the Guhathakurta
et al. (2006) results, reaffirms the existence of this metallicity
gradient.

\begin{figure*}[t]
\epsscale{1.1}
\plotone{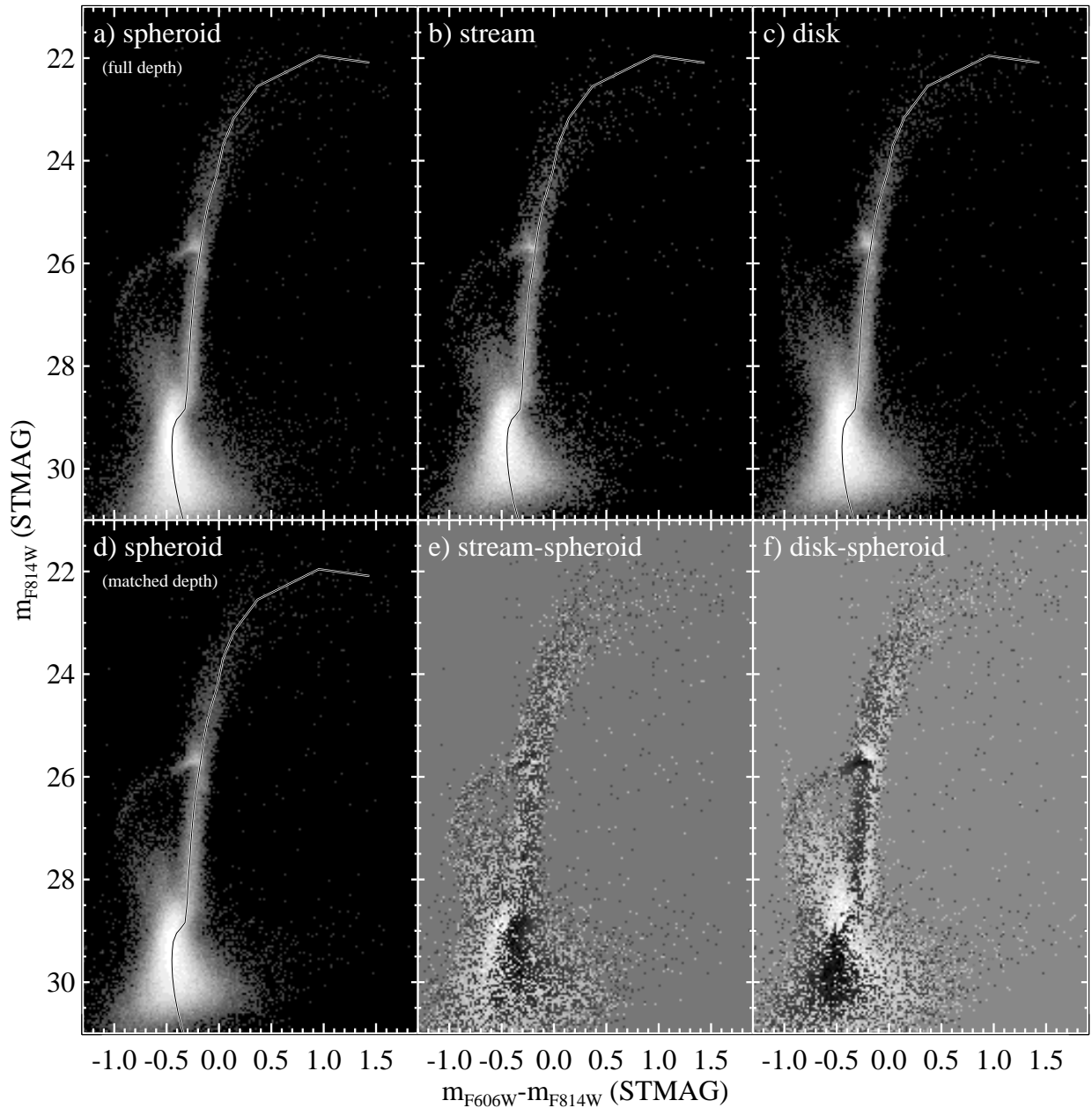}
\epsscale{1.0}
\caption{Comparisons of the CMDs for our three fields.  The ridge line
for NGC~104 ({\it curve}) is shown for reference.  {\it a)} The CMD of the
spheroid field, shown at its full depth, at a logarithmic stretch.
{\it b)} The CMD of the stream field.  {\it c)} The CMD of the disk
field.  {\it d)} The CMD of the spheroid field, shown at a depth that
matches that in the stream and disk.  {\it e)} The difference between
the stream and spheroid CMDs (with the latter scaled to match the
number of stars in the former).  The RGB and HB distributions are very
similar, but the locus of stars at the MS extends slightly brighter and
bluer than that in the spheroid.  {\it f)} The difference between
the disk and spheroid CMDs (with the latter scaled to match the number
of stars in the former).  The RGB of the disk is considerably redder
than that of the spheroid, indicating higher metallicities in the
disk.  The HB of the disk is almost entirely in the red clump, with a
spread to brighter luminosities, indicating higher metallicities and
younger ages in the disk.  The blue plume of stars above the MSTO is
much stronger in the disk, indicating younger ages than in the
spheroid.}
\label{compare}
\end{figure*}

Although the CMDs for each field have many similarities, closer
inspection reveals significant distinctions, especially between the
disk and the other two fields.  We highlight these distinctions in
Figure~\ref{compare}, which shows the differences between the stream
and spheroid and also those between the disk and spheroid.  The
spheroid data used in each comparison are a subset that reaches
approximately the same depth as the stream and disk data.  The
spheroid CMD was also scaled to the number of stars in each of the
other two CMDs before subtracting; note that it makes little
difference if this normalization is done based on the total number of
stars in each field or just those well above the detection limits
(e.g., $m_{F814} < 28$~mag).  Relative to the spheroid main sequence,
the stream main sequence extends somewhat farther to the blue, even
though the RGB and HB distributions are nearly identical.  Thus, the
age distribution in the stream must extend to slightly younger ages
than those in the spheroid (as also noted by Brown et al.\ 2006). In
contrast, the distributions of age and metallicity in the disk extend
to significantly younger ages and higher metallicities than those in
the spheroid and stream, and the old metal-poor population is not as
prominent. The RGB stars in the disk are skewed toward redder colors,
while the HB population is largely restricted to the red clump; both
of these features indicate a higher metallicity in the disk.  In the
disk population, the red clump HB is also somewhat extended in
luminosity, indicating a younger age distribution (an excellent
example of the variation in clump luminosity with age can be seen in
the Monelli et al.\ [2003] study of the Carina dwarf spheroidal).
There does not appear to be a significant population on the blue HB,
although a trace population might be hidden in the blue plume of stars
rising above the dominant MSTO; Figure~\ref{compare}f shows an
oversubtraction of the blue HB from the spheroid ({\it dark boxes})
appearing within the cloud of undersubtracted blue plume stars from
the disk ({\it light boxes}).  The stronger plume in the disk
population indicates an extension to significantly younger ages.  The
plume in the disk population includes $\approx$40 stars that are
brighter than the region where the blue end of the HB would nominally
fall, implying that these bright blue stars have masses of
$\sim$2--5~$M_\odot$ and ages of $\sim$0.2--1 Gyr.  Note that
Cuillandre et al.\ (2001) have also seen evidence for trace
populations of young stars in the outer disk of M31.  However, the
disk does not look quite as young as one might expect if there were a
significant thin disk population -- a point we will return to later.

In a field population, it is difficult to distinguish between young
metal-poor stars and old blue stragglers (see Carney, Latham, \& Laird
2005 and references therein).  Thus, some of the apparently young
stars in our CMDs ($\lesssim 6$~Gyr) might instead be blue stragglers.
However, whether blue stragglers form via merger or mass transfer, in
an old population they will be limited to $M \lesssim $2~$M_\odot$.
All three of our fields show blue plume stars as bright as the HB over
a wide range of color, and in the disk these stars continue to
luminosities significantly brighter than the HB.  The high masses
required to explain the brightest stars in the blue plume population
imply that truly young stars are present, and these stars appear to be
a smooth extension of the fainter blue plume population.  This argues
against a significant contribution from blue stragglers in the blue plume.

\begin{figure*}[t]
\epsscale{1.1}
\plotone{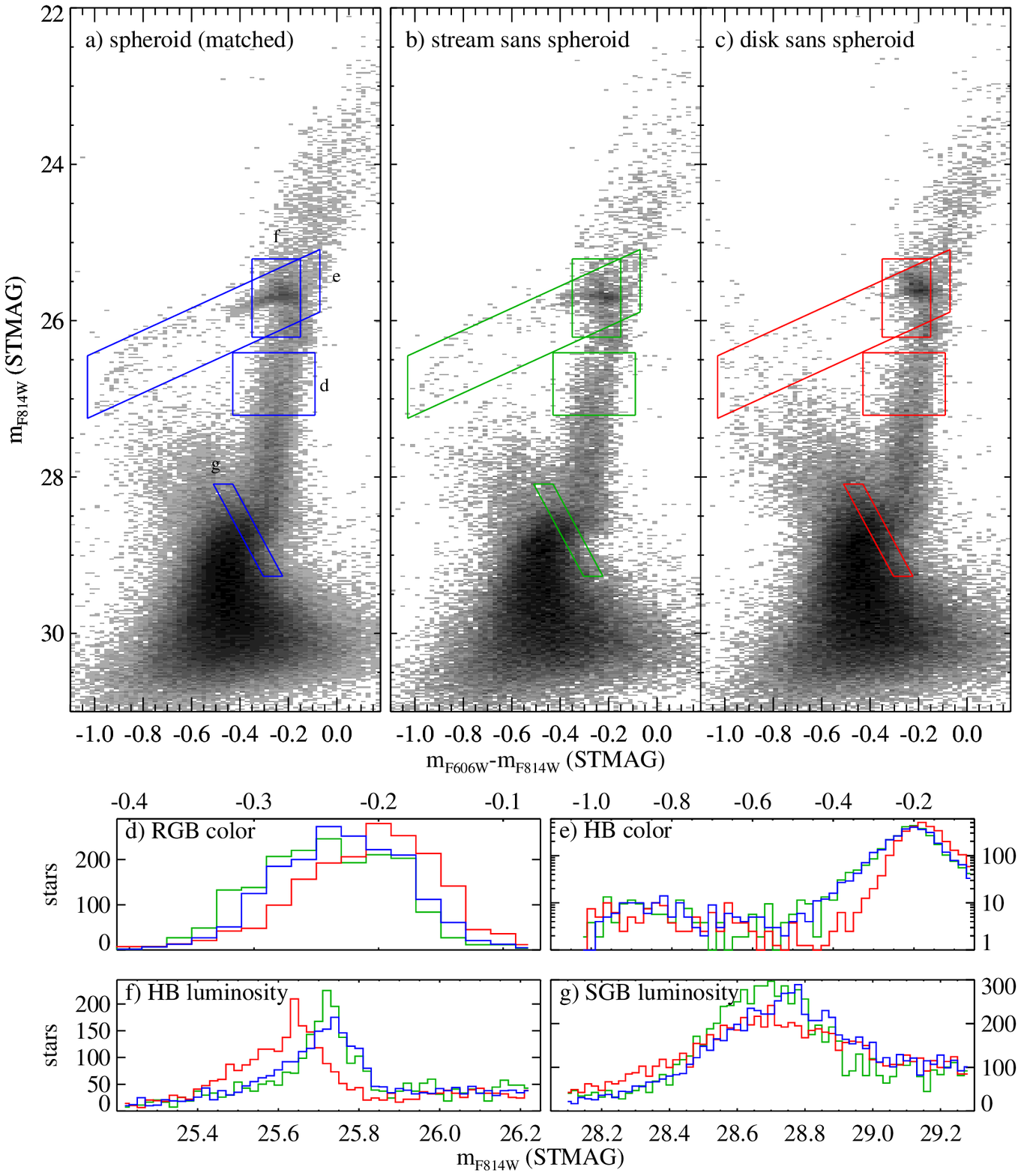}
\epsscale{1.0}
\caption{The spheroid CMD compared to the stream and disk CMDs, where
we have attempted to subtract the spheroid contamination 
from the stream and disk.
{\it a)} The CMD of the spheroid field, shown at a depth that
matches that in the stream and disk fields.  Cuts across the CMD ({\it
blue boxes}) are used to make comparisons with the stream and disk;
the histograms in each cut ({\it panels d-g}) are normalized to the
number of stars in the spheroid.  Labels refer to subsequent panels in
this figure.  {\it b)} The CMD of the stream field, with a subtraction
of spheroid stars assumed to contaminate at the 25\% level, and with
the same cuts indicated ({\it green boxes}).  {\it c)} The CMD of the
disk field, with a subtraction of spheroid stars assumed to
contaminate at the 33\% level, and with the same cuts indicated ({\it
red boxes}).  {\it d)} Histograms for stars along the RGB color cut
for the spheroid ({\it blue}), stream ({\it green}) and disk ({\it
red}). {\it e)} Histograms along the HB color cut.  {\it f)}
Histograms along the HB luminosity cut.  {\it g)} Histograms along the
SGB luminosity cut.  Compared to the spheroid and stream, the disk
population has a redder RGB (indicating higher metallicities), an HB
that falls mostly in a red clump that extends to brighter luminosities
(indicating younger ages and higher metallicities), and a stronger
blue plume above the MSTO (indicating younger ages).}
\label{subtract}
\end{figure*}

If we fit Gaussian distributions to the velocity data in our fields
(Figure~\ref{cmdsvels}), we find that the spheroid is a $\sim$25\%
contamination in our stream field and a $\sim$33\% contamination in
our disk field.  Given the wide separation between our fields
(Figure~\ref{mapfig}), we cannot necessarily assume that the
population in our spheroid field is representative of the spheroid
contamination in our stream and disk fields.  However, it is natural
to ask how the stream and disk CMDs would look if the spheroid
contamination were subtracted under the assumption that the population
in our spheroid field is in fact representative of this contamination.
To show this, we used that subset of the spheroid data that is matched
to the depth of the stream and disk data.  We randomly drew a star
from these spheroid data, found the star in the stream data that most
closely agreed in its photometry, and then subtracted that star from
the stream data.  These subtractions were repeated until 25\% of the
stream stars were removed.  In 99\% of the subtractions, the star
subtracted from the stream data was within 0.02~mag of the spheroid
star, and in 99.9\% of the subtractions, the star subtracted from the
stream data was within 0.1~mag of the spheroid star; the handful of
stars that could not be matched at this level fell very far from the
dominant stellar locus (in the negligible cloud of sparse stars at
random colors and magnitudes), and these were not subtracted.  We
repeated this process on the disk data, but there subtracted 33\% of
the disk stars; again, 99\% of the subtractions matched disk to
spheroid stars within 0.02~mag, while 99.9\% of the subtractions
matched disk to spheroid stars within 0.1~mag.  The resulting CMDs are
shown in Figure~\ref{subtract}.  Because of the many similarities
between the original three CMDs (Figure~\ref{cmdsvels}), the changes
due to the subtraction of the spheroid contamination are subtle.  To
help highlight the differences between the three fields, we also show
luminosity and color cuts across the CMDs ({\it colored boxes});
panels $d$ and $e$ show the color distributions on the lower RGB and
HB, respectively, while panels $f$ and $g$ show the luminosity
distributions at the red clump and SGB, respectively.  The color and
luminosity cuts help quantify the similarities and differences between
the populations discussed above and shown in Figure~\ref{compare}.
Compared to the spheroid population, the stream population exhibits
similar RGB and HB morphologies, but its main sequence extends
somewhat brighter and bluer.  In contrast, the disk population
exhibits RGB and HB morphologies that are skewed toward redder colors,
with the main sequence showing a strong extension to brighter and
bluer colors.

\subsection{Maximum Likelihood Fitting of Isochrones}

We turn now to the quantitative fitting of our CMDs.
Our characterization of the star formation history in each field
primarily uses the StarFish code of Harris \& Zaritsky (2001).  This
code takes a grid of isochrones, populates them according to the 
initial mass function (IMF),
then applies the photometric scatter and incompleteness (as a function
of magnitude and color) determined in the artificial star tests.  The
code then fits the observed CMD by employing linear combinations of the
scattered isochrones.  The fitting can be done via minimization of
either a $\chi^2$ statistic or the Maximum Likelihood statistic of
Dolphin (2002).  We found little difference between fits done with
either statistic, and ultimately used the Maximum Likelihood statistic
in our analysis.

In the StarFish fitting, each isochrone at a given age and metallicity
is varied independently, resulting in a large number of free
parameters in the fit.  This method is similar to most of the star
formation history methods used in the literature (e.g., Dolphin 2002;
Skillman et al.\ 2003).  Although the term ``star formation history''
might imply a physical connection between the subpopulations, this
method is really a fit to the age and metallicity distributions.  In
addition to StarFish, we wrote our own codes that fit the isochrones to
the data according to mathematical and physical restrictions that
greatly reduce the number of free parameters; these models will be 
the subject of a future paper.

We do not fit the entire range of stars observed in the CMD.  Instead,
we restrict our fits to the lower RGB (below the level of the HB),
SGB, and upper main sequence.  Specifically, we fit over $-0.9 \leq
m_{F606W}-m_{F814W} \leq -0.1$~mag in color, and $26.5 \leq m_{F814W}
\leq 30.5$~mag in magnitude for the spheroid data and $26.5 \leq
m_{F814W} \leq 30.0$~mag in magnitude for the stream and disk data
(which are $\approx$0.5~mag shallower).  This region of the CMD
offers excellent sensitivity to age and metallicity while avoiding
those regions of the CMD that have low signal-to-noise ratio or that are poorly
constrained by the models (such as the HB, the upper RGB, the RGB
luminosity function bump, and the faint end of the CMD).  The HB is a
qualitative indicator of age and metallicity, becoming
redder at younger ages and higher metallicities, and eventually forming a
red clump with a significant spread in luminosity.  However,
disentangling the effects of age and metallicity is highly uncertain;
indeed, the ``second parameter debate'' in the study of HB morphology
refers to the dependence of the HB morphology on parameters other than
metallicity, such as age and helium abundance.  Although Galactic
foreground stars comprise much less than 1\% of the stars in our
field, they tend to fall near the upper RGB in M31, which is sparsely
populated in our data; the upper RGB is thus the one region of our
CMDs with significant foreground dwarf contamination.  In addition,
the upper RGB is contaminated by asymptotic giant branch (AGB) stars,
which in turn have a distribution depending on the age and [Fe/H] of
their progenitor HB stars.  The RGB luminosity function bump is a
qualitative metallicity indicator, and it is most prominent in CMDs of
metal-rich populations, where it appears as an overdensity on the RGB
immediately below the luminosity of the HB; theoretical models
reproduce the general trend for the bump luminosity to brighten with
decreasing metallicity, but the zeropoint of the relationship is
uncertain, and the mix of age and metallicity in our populations makes
it difficult to interpret this feature in the data.  The faintest main
sequence stars in the CMD suffer from large photometric scatter and
low completeness.

We use the Victoria-Regina Isochrones (VandenBerg et al.\ 2006) in all
of our fitting.  These isochrones do not include core He diffusion,
which would decrease their ages at a given turnoff luminosity by $\sim
10$\% (VandenBerg et al.\ 2002).  Although the ages of isochrones with
core He diffusion are likely more accurate, models in which diffusion
is allowed to act efficiently on other elements in the surface layers
show significant discrepancies when compared to observed CMDs,
indicating that there must be some other mechanism at work, such as
turbulence in the surface layers (see Brown et al.\ 2005 and
references therein).  Helium diffusion can still occur in the core,
and thus the ages discussed herein should be reduced by $\sim$10\% to
obtain absolute ages.

The Victoria-Regina Isochrones are distributed with a ground-based
magnitude system.  Sirianni et al.\ (2005) provide an iterative
transformation to put ACS data in a ground-based system, but warn
against its use, given the systematic errors intrinsic to such a
process.  The biggest problem is that the F606W bandpass is very
different from Johnson $V$, although the difference between F814W and
Cousins $I$ is nonnegligible, too.  To properly make the
transformation from one system to the other, one must know the
intrinsic spectral energy distribution of the source, and this is
difficult to estimate based on photometry in two broad bandpasses.
It is much more straightforward to use the physical parameters along
each model isochrone (effective temperature and surface gravity) to
transform the models into the observational system using synthetic
spectra of the appropriate metallicity.  We use the transformation of
Brown et al.\ (2005), which produces good agreement between these
isochrones and the ACS observations of Galactic clusters spanning a
wide range in metallicity (Table~\ref{tabclus}).  Over most of the CMD
(including the region we use here for fitting), the agreement is
better than $\sim$0.02~mag.  In this sense, we are using the
isochrones to provide relative changes in age and metallicity, once
they have been anchored to observations of Galactic clusters.  We are
thus providing star formation histories in a reference frame based on
the ages and metallicities of the clusters listed in
Table~\ref{tabclus}.

\subsubsection{The Isochrone Grid}

\begin{figure}[h]
\epsscale{1.2}
\plotone{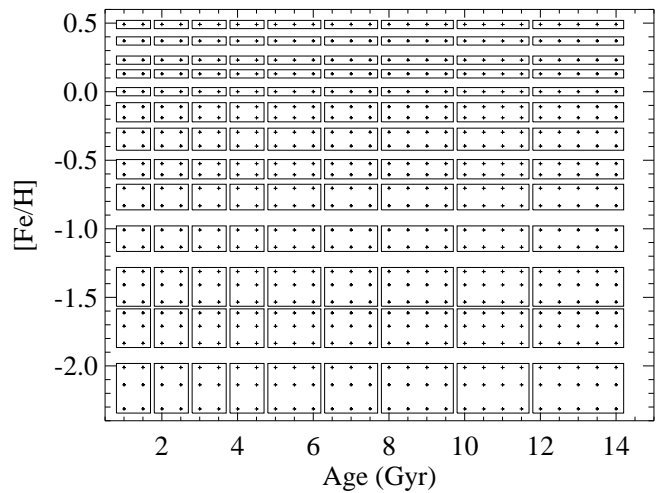}
\epsscale{1.0}
\vskip -0.1in
\caption{The isochrones used in StarFish fitting.  A fine grid of 
isochrones ({\it crosses}) was used to avoid artificial lumpiness
in the synthetic CMD, but these isochrones were locked together in
groups ({\it boxes}) to reduce the number of free parameters and
to avoid degeneracies in the fit.}
\label{gridfig}
\end{figure}

We fit a large grid of isochrones spanning $1 \leq$~age~$\leq 14$~Gyr
(with 0.5 Gyr steps) and $-2.3 \leq $~[Fe/H]~$\leq +0.5$ (with
$\approx$0.1 dex steps) using the StarFish code.  The fine spacing in
age and metallicity avoids artificial lumpiness in the synthetic
CMDs but means that neighboring isochrones in the grid are nearly
degenerate.  Such degeneracies, plus the large number of free
parameters, do not allow a fit to converge in a reasonable time.
Fortunately, the StarFish code allows groups of neighboring isochrones
to be locked such that their amplitudes vary together; one of these
isochrone groups is treated as a single isochrone as far as the
fitting is concerned, even if its stars span a small range in age and
metallicity (see Harris \& Zaritsky 2001 for details).  We locked our
full grid of isochrones into 117 independent isochrone groups, with
the sampling chosen to match the nonlinear changes in the CMD with age
and metallicity (the CMD changes more rapidly at higher metallicities
and younger ages).  The grid of isochrones and the locked isochrone
groups are shown in Figure~\ref{gridfig}.

\subsubsection{Fixed Parameters}

Besides distance and reddening, there are several other parameters
that must be fixed before proceeding with a fit.  The binary fraction
is highly uncertain, and not even well-constrained in the field or
cluster populations of our own Galaxy; the value appears to be in the
range of 10--30\% in the field population of the Galactic halo (Ryan
1992 and references therein).  The fits to our data are best when the
binary fraction is near 10\%, whereas fits with the binary fraction
significantly deviating from 10\% show noticeable residuals.  Thus,
unless specified otherwise, the binary fraction is assumed to be 10\%
throughout this paper.  The binary fraction is set in the StarFish
code (Harris \& Zaritsky 2001) at the stage where it scatters the
isochrones; specifically, for a given fraction of stars, it draws a
second star randomly from the IMF and produces a single unresolved
object with the combined color and magnitude of the two stars.  For
the IMF index, we chose the Salpeter (1955)
value of $-1.35$.  For the isochrone abundances, we did not assume a
scaled-solar abundance pattern.  Instead, we assumed that the alpha
elements are enhanced at low metallicity and unenhanced (scaled-solar)
at high metallicity; specifically, we assumed [$\alpha$/Fe]~=~0.3 at
[Fe/H]~$\leq -0.7$ and [$\alpha$/Fe]~=~0.0 at [Fe/H]~$> -0.7$.  At the
[$\alpha$/Fe] resolution available in our isochrone grid, this trend
roughly reproduces that seen in the Galaxy (Pritzl, Venn, \& Irwin
2005 and references therein), although bulge populations appear to be
enhanced in alpha elements even at high metallicity (McWilliam \& Rich
1994; Rich \& Origlia 2005).  As it turns out, the IMF and
alpha-enhancement assumptions make little difference in our results.
All of these assumptions (distance, reddening, binary fraction, IMF,
and alpha enhancement) are varied in our exploration of systematic
errors (see \S\ref{syssec}).

\subsubsection{Uncertainties}
 
In the fits below, we do not plot error bars for the weights of the
individual isochrones.  This is because the uncertainty associated
with the normalization of any individual isochrone is very large, and
correlated with the normalization of neighboring isochrones.  If any
one isochrone in the best-fit model is deleted from the fit,
compensating changes can be made in neighboring
isochrones that restore the quality of the fit.  The result is that
the uncertainty on any individual isochrone weight is largely
meaningless. These difficulties are a continuing plague for studies of
star formation histories in complex populations (e.g., Skillman et
al.\ 2003; Harris \& Zaritsky 2004).  If one is fitting a simple
stellar population (single age and single metallicity), one can trace
out confidence contours in the age-metallicity plane according to the
change in fit quality, but with a complex star formation history, it
is the distribution of ages and metallicities that matters.  What one
really wants is a set of isochrones that are truly eigenfunctions of
an orthogonal basis set.  However, there is not an obvious basis
function that relates in a simple way back to physical parameters.
The sampling in our isochrone grid is fine enough to avoid artificial
structure in the synthetic CMDs, yet coarse enough to avoid isochrones
that are completely degenerate within the photometric errors.

Even though some of the isochrone weights in the final fits are very
small, the ensemble of these small weights is necessary for a good
fit.  One way of demonstrating this assertion is by repeating the fits
after deleting isochrones with low weights.  Starting with the best
fit, we first sorted the isochrones by their fitted weights, and then
retained only those whose weight exceeded a specified cutoff;
specifically, the cutoff in weight was chosen so that this subset of
isochrones accounted for 90\% of the stars in the best fit.  Refitting
with this reduced set of isochrones produced terrible fits (fit score
$\sim$50\% larger).  The fit was also poor when we retained those
isochrones responsible for 95\% of the stars in the best fit.  The fit
did not become acceptable until we had retained those isochrones
responsible for 99\% of the stars in the best fit ($\sim$50 of the
original 117 isochrone groups).

\subsection{Results for the Spheroid}
\label{secsph}

\begin{figure}[h]
\epsscale{1.2}
\plotone{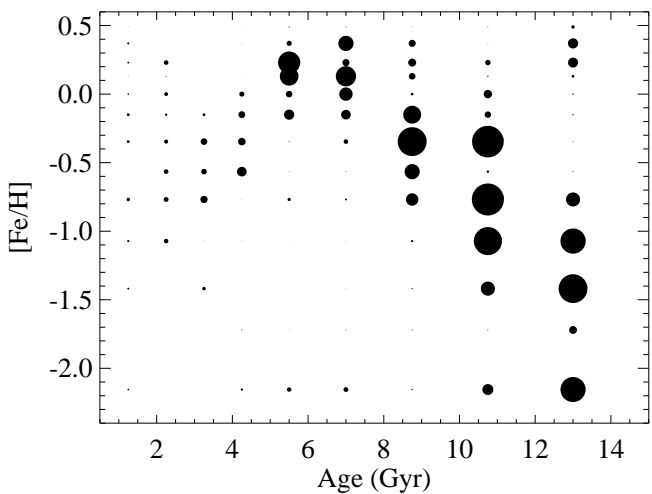}
\epsscale{1.0}
\vskip -0.1in
\caption{The distribution of age and metallicity in the best-fit model
of the spheroid data.  The area of the filled circles is proportional
to the number of stars in each isochrone group.
}
\label{halofit}
\end{figure}

The distribution of age and metallicity in our best fit to the
spheroid data is shown in Figure~\ref{halofit}.  In this figure, the
area of the symbols ({\it filled circles}) is proportional to the
number of stars in each isochrone group.  Note that the spacing of the
isochrone groups is irregular, so that if one were to plot a star
formation rate in units of $M_\odot$ per unit time per unit
logarithmic metallicity, the relative sizes of the symbols would be
somewhat increased at younger ages and higher metallicities (where the
spacing is finer).  As noted by Brown et al.\ (2003), the
spheroid CMD is best fitted by a wide range of age and metallicity, and
is strikingly different from the old, metal-poor halo of the Milky
Way.  Approximately 40\% of the stars are less than 10~Gyr old, and
approximately 50\% of the stars are more metal-rich than 47~Tuc
([Fe/H]$\approx-0.7$).  The mean metallicity, $<$[Fe/H]$>$=$-0.6$, is
identical to that found by Durrell et al.\ (1994) at 9~kpc on the
minor axis, and slighter higher than the $<$[m/H]$>$=$-0.6$ found by
Holland et al.\ (1996) from earlier WFPC2 photometry of our field,
with similar spreads to both higher and lower metallicities.  Although
our mean metallicity is much higher than that in the Milky Way halo,
the metallicity distribution definitely has a tail extending to
metal-poor stars.  These include the RR Lyrae stars in our field,
which have a mean metallicity of [Fe/H]~=~$-1.7$ (Brown et al.\
2004a), and the minority population of blue HB stars.

Although we used the Dolphin (2002) Maximum Likelihood statistic to
perform our fits, we also compared the results with those obtained
from a traditional $\chi^2$ statistic, and the fits were
similar.  Dolphin (2002) also provides a goodness of fit statistic,
$\chi^2_{\rm eff}$, for those more familiar
with $\chi^2$ fitting (with values close to unity indicating a good
fit).  The best fit model (Figure~\ref{halofit}) has $\chi^2_{\rm eff}
=1.11$ per degree of freedom (8000 CMD bins minus 117 freely varying
isochrone weights).  This score clearly implies an imperfect fit.  To
demonstrate this, we ran Monte Carlo simulations of the idealized
case.  We created random realizations of the data drawn from the
best-fit model to obtain the distribution of the Maximum Likelihood
statistic, and found that the Maximum Likelihood statistic obtained in
our best-fit model exceeds the mean score by 6$\sigma$ (where $\sigma$ is
one standard deviation in the distribution of the Maximum Likelihood
statistic from the Monte Carlo runs).

\begin{figure}[h]
\epsscale{1.2}
\plotone{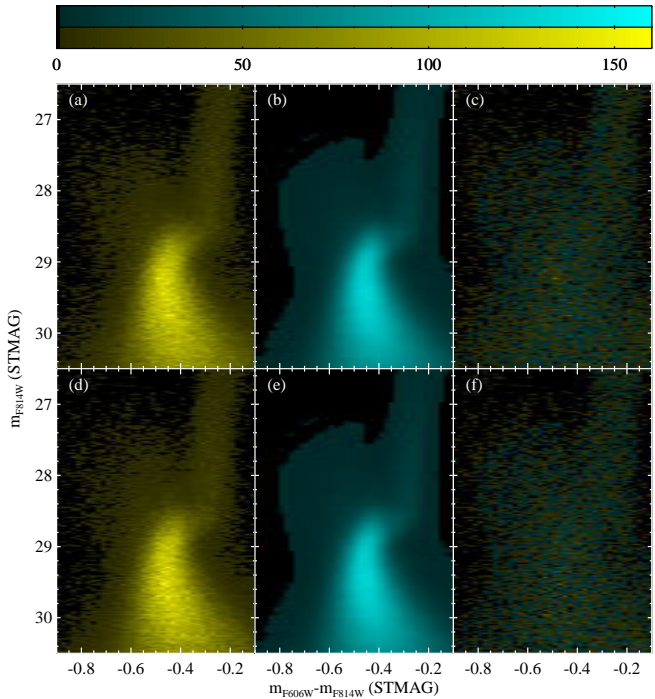}
\epsscale{1.0}
\vskip -0.1in
\caption{{\it Top panels:} The CMD of the spheroid data ({\it
yellow}), the best-fit model to those data ({\it blue}), and the
differences between the data and model ({\it yellow} and {\it blue}),
all shown at the same linear stretch.  {\it Bottom panels:} An
artificial CMD drawn from the best-fit model ({\it yellow}), the same
best-fit model ({\it blue}), and the differences between the
artificial data and model ({\it yellow} and {\it blue}), all shown at
the same linear stretch employed in the top panels.  }
\label{halomod}
\end{figure}

There are many reasons why the model should not
exactly reproduce the data.  These include imperfections in the
isochrones (they are calibrated at the $\sim$0.02~mag level against
Galactic globular clusters observed in the same filters), deviations
from a Salpeter (1955) IMF, deviation from our assumed binary fraction
of 10\% (e.g., one might imagine that the binary fraction varies with
age and metallicity depending on the variations in the formation
environment), and the limitations of the artificial star tests used to
scatter the isochrones (artificial stars are created, with noise, from
the same PSF model used in the PSF fitting, while real stars will
deviate from the PSF model due to noise and true intrinsic
inaccuracies in the PSF model).  Although the model does not exactly
reproduce the data distribution over 8000 CMD bins, the deviations are
remarkably small, as we show in Figure~\ref{halomod}.  In the top row
of panels, we show the data in the fitting region ({\it yellow}), the
best-fit model ({\it blue}), and the differences between the two ({\it
yellow} and {\it blue}) shown at the same linear stretch; i.e., 
the CMD bins in panel $c$ are shaded blue where the model exceeds the
data, and shaded yellow where the data exceeds the model, with the
shading on the same linear scale employed in panels $a$ and $b$.  The
differences between the data and model appear almost completely
random, with minimal systematic residuals; in fact, the upper panels look
much like the idealized case shown in the bottom row of panels, where
the residuals are completely random.
There, we show a random realization of the best-fit model ({\it
yellow}), a repeat of the best-fit model ({\it blue}), and the
differences between the two ({\it yellow} and {\it blue}).  The
realization ({\it bottom left}) is nearly indistinguishable from the
actual data ({\it top left}).  The difference between the realization
and the model ({\it bottom right}) demonstrates the noise residuals one can
expect when comparing a smooth model to discrete data in the idealized
case ($\chi^2_{\rm eff} = 1$).  

\begin{figure}[h]
\epsscale{1.2}
\plotone{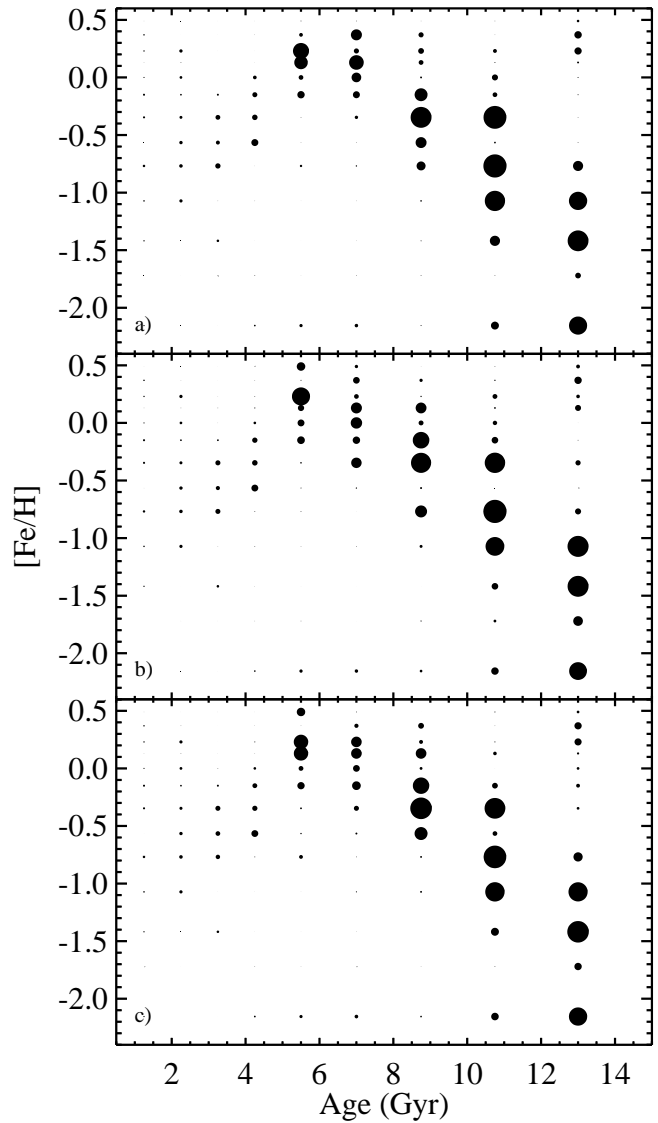}
\epsscale{1.0}
\vskip -0.1in
\caption{Three different attempts at fitting the spheroid data, using
the same isochrones in each case, but where the initial guess in each
panel was a distinct random distribution of isochrones.
The area of the filled circles is proportional to the
number of stars in each isochrone group.  Although there are small
variations in the individual amplitude weights from panel to panel, it
is clear that the best-fit model is well converged.}
\label{converge}
\end{figure}

Given the large number of free parameters, one might also wonder if
the ``best-fit'' model has truly converged on the best fit.  One way
to test this is through repeated fitting with distinct initial
conditions.  We show in Figure~\ref{converge} the results of three
``best-fit'' models to the spheroid data, each of which started from a
distinct random set of isochrone weights.  Although there are small
variations in the final individual isochrone weights, it is clear that
the overarching result is the same in each case.  As stated earlier,
the degeneracies in the isochrone set mean that any individual
isochrone can be varied significantly without changing the fit
quality.  For example, in Figure~\ref{converge}, the relatively low
weight at [Fe/H]=$-1.7$, compared to the weights at [Fe/H]=$-1.4$
and [Fe/H]=$-2.1$, is not meaningful; for the isochrones at 13~Gyr,
we can redistribute the weights at [Fe/H]=$-1.4$, $-1.7$, and $-2.1$
so that they are the same in each of these bins, and the fit quality
does not suffer.

\begin{figure}[h]
\epsscale{1.2}
\plotone{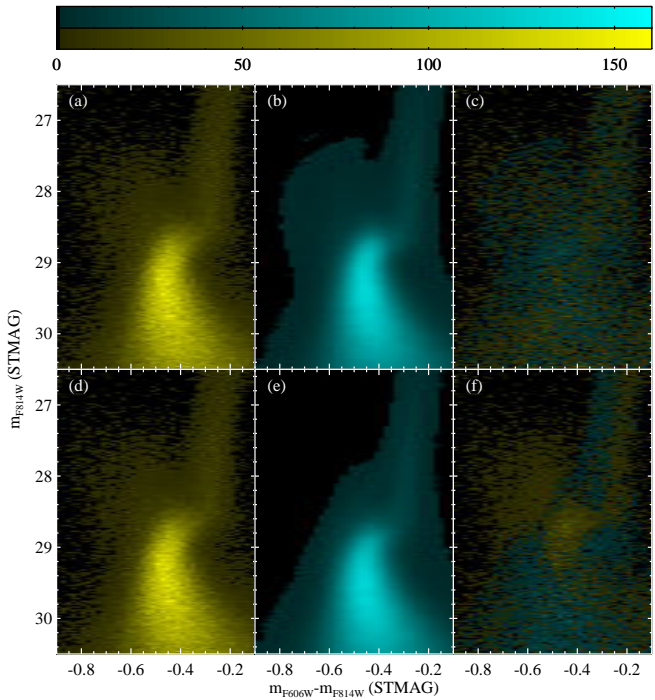}
\epsscale{1.0}
\vskip -0.1in
\caption{{\it Top panels:} The CMD of the spheroid data ({\it
yellow}), the best-fit model to those data using a set of isochrones
restricted to ages less than 10~Gyr ({\it blue}), and the
differences between the data and model ({\it yellow} and {\it blue}),
all shown at the same linear stretch.  {\it Bottom panels:} The same
CMD of the spheroid data ({\it yellow}), the best-fit model to those
data using a set of isochrones restricted to ages $\ge 10$~Gyr
({\it blue}), and the differences between the data and
model ({\it yellow} and {\it blue}), all shown at the same linear
stretch employed in the top panels.  It is clear that neither model
is acceptable, given the residuals ({\it right hand panels}). }
\label{haloyngold}
\end{figure}

Although the uncertainties on the individual isochrone weights in the
best-fit model are large, one can ask what classes of models, in a
broad sense, produce fits that are as good as the best-fit model.  If
one restricts the fit to isochrones of ages$<10$~Gyr, the quality of
the fit is noticeably reduced, with $\chi^2_{\rm eff} = 1.18$ (a fit that
is an additional 5$\sigma$ worse than the best-fit model).  Much of
the weight in this fit falls at the top end of the allowed age range,
and the difference between the model CMD and the data CMD shows
significant residuals (Figure~\ref{haloyngold}).  Alternatively, if
one restricts the fit to isochrones with ages$\ge 10$~Gyr, the quality
of the fit is grossly reduced, with $\chi^2_{\rm eff} = 3.09$ and
very obvious differences between the model CMD and the data CMD
(Figure~\ref{haloyngold}).  This is consistent with the results of
Brown et al.\ (2003), who showed that the spheroid CMD is inconsistent
with a purely old population of stars.

\begin{table*}[t]
\begin{center}
\caption{Summary of Spheroid Fitting}
\begin{tabular}{lcccl}
\tableline
Model & $<$[Fe/H]$>$ & $<$age$>$ & $\chi^2_{\rm eff}$ & Comment \\
\tableline
Standard model & $-0.6$ & 9.7 & 1.11 & Minimal residuals in fit\\
Age $< 10$~Gyr & $-0.5$ & 8.4 & 1.18 & Significant residuals in fit\\
Age $\ge 10$~Gyr & $-0.8$ & 10.9 & 3.09 & Gross residuals in fit\\
No old metal-rich stars & $-0.6$ & 9.6 & 1.11 & Minimal residuals in fit\\
No young metal-poor stars & $-0.6$ & 9.7 & 1.18 & Misses part of plume\\
\tableline
\end{tabular}
\label{tabspheroid}
\end{center}
\end{table*}

The best-fit model has minority populations in the isochrones
representing old metal-rich stars and young metal-poor stars.  If
truly present, these populations are extremely interesting, because
the former imply that at least some of the stars were formed in
something like a bulge environment (with rapid early enrichment),
while the latter imply the accretion of metal-poor stars from dwarf
galaxies or star formation following the infall of
relatively pristine material.  To test this, we repeated the fit while
excluding two regions from the input grid of isochrones: age$\ge
10$~Gyr at [Fe/H]$\ge 0$, and age$< 5$~Gyr at [Fe/H]$< -0.5$; each of
these regions contains 3\% of the stellar mass in the best-fit model.
If the old metal-rich isochrones are excluded from the fit, the fit
quality in the resulting model does not suffer at all; thus, the CMD
is consistent with either a small population of such old metal-rich
stars or none at all.  In contrast, if the young metal-poor isochrones
are excluded from the fit, the fit quality is somewhat reduced, with
$\chi^2_{\rm eff} = 1.18$, due to the model missing the brightest and
bluest stars in the blue plume above the dominant main sequence.  
This is not surprising, given our visual inspection of the
CMD and comparison to young isochrones of various metallicities
(Figure~\ref{comparegrid}).  Note that the scattered model isochrones
include the effects of blends (determined by the artificial star tests)
but not any contribution from blue stragglers; thus, some (but not all) of the
young stars in the fit ($\lesssim 6$~Gyr) could be an attempt to
account for blue stragglers (see \S\ref{secinspec}).

We summarize the fits to the spheroid data in Table~\ref{tabspheroid}.
Our standard model is that which simply allows the full grid
(Figure~\ref{gridfig}) to vary freely, while the other models are
self-explanatory.  Mean values of [Fe/H] and age are not as useful as
the full age and metallicity distributions, given the complicated star
formation history present in the field, but these mean values do serve
as a yardstick to gauge differences between the fits. \\

\subsection{Results for the Stream}

\begin{figure}[h]
\epsscale{1.2}
\plotone{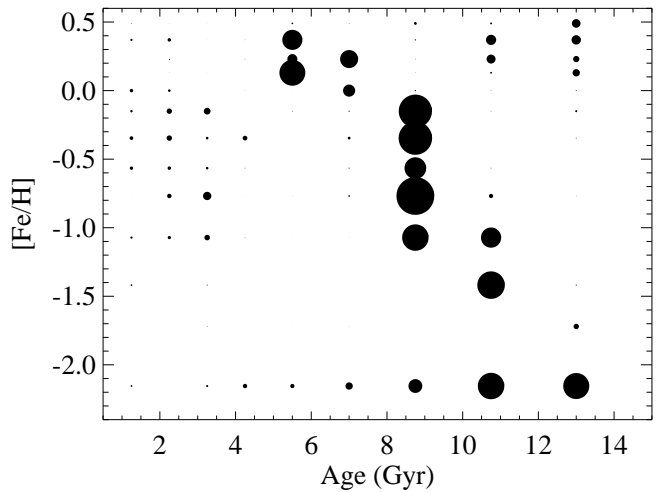}
\epsscale{1.0}
\vskip -0.1in
\caption{The distribution of age and metallicity in the best-fit model
of the stream data.  The area of the filled circles is proportional to
the number of stars in each isochrone group.  The total area within
the filled symbols has been normalized to that in
Figure~\ref{halofit}, to ease comparison (but in reality the surface
brightness in the stream is $\sim$0.5~mag fainter).  }
\label{strmfit}
\end{figure}

\begin{figure}[h]
\epsscale{1.2}
\plotone{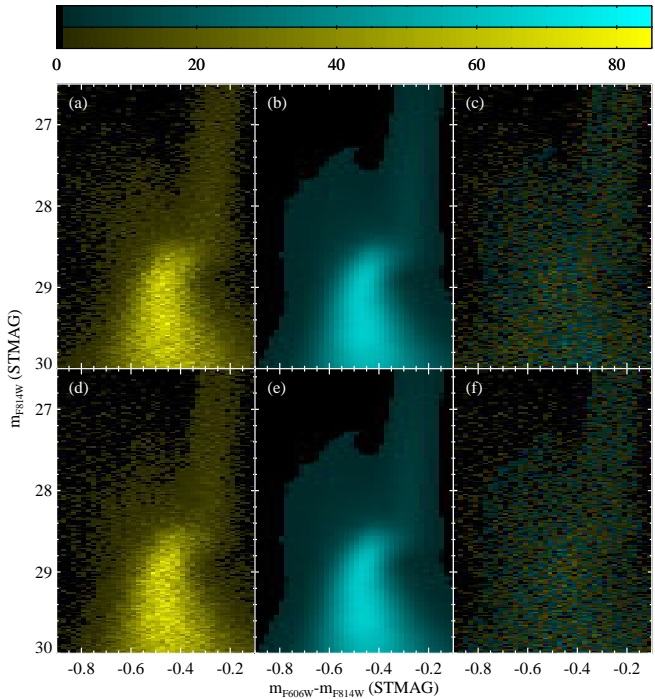}
\epsscale{1.0}
\vskip -0.1in
\caption{{\it Top panels:} The CMD of the stream data ({\it yellow}),
the best-fit model to those data ({\it blue}), and the differences
between the data and model ({\it yellow} and {\it blue}), all shown at
the same linear stretch.  {\it Bottom panels:} An artificial CMD drawn
from the best-fit model ({\it yellow}), the same best-fit model ({\it
blue}), and the differences between the artificial data and model
({\it yellow} and {\it blue}), all shown at the same linear stretch
employed in the top panels.  Note that the magnitude range of the
stream fit is smaller than that in the spheroid fit, because the
spheroid data are $\sim$0.5 mag deeper than the stream data.}
\label{strmmod}
\end{figure}

The distribution of age and metallicity in our best fit to the
stream data is shown in Figure~\ref{strmfit}.  Given the qualitative
similarities between the stream and spheroid CMDs, it is not
surprising that the best-fit distribution of age and metallicity in
the stream resembles that in the spheroid.  However, as noted above,
there are some distinctions.  The mean age in the stream (8.8~Gyr) is
$\sim$1~Gyr younger than that in the spheroid (9.7~Gyr), while the
mean metallicities are nearly the same ($-0.6$ in the spheroid and
$-0.7$ in the stream).  The fit quality for the best-fit stream model
is similar to that for the spheroid, with $\chi^2_{\rm eff} = 1.08$.  
In Figure~\ref{strmmod}, we show the comparison of the best-fit
model to the data, as well as the residuals.

\begin{figure}[h]
\epsscale{1.2}
\plotone{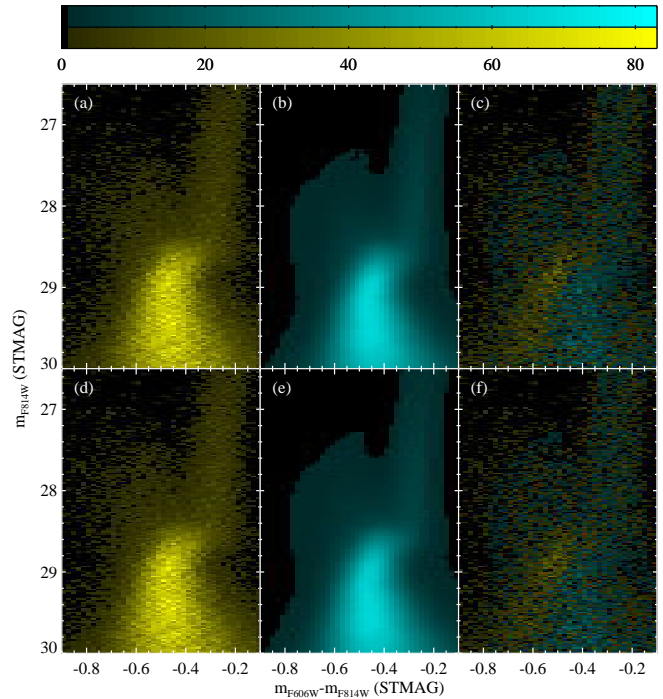}
\epsscale{1.0}
\vskip -0.1in
\caption{{\it a)} The CMD of the stream data ({\it yellow}).  {\it b)}
A model for the stream data ({\it blue}), but constructed from the
spheroid star formation history, scattered with the observational
errors of the stream data and normalized to the stream star counts. {\it c)}
The differences between the data and model ({\it
yellow} and {\it blue}), all shown at the same linear stretch.  {\it
d)} The same CMD of the stream data ({\it yellow}). {\it e)} The
best-fit compromise model fit simultaneously to the spheroid and
stream datasets ({\it blue}).  {\it f)} The differences between the data and
model ({\it yellow} and {\it blue}), all shown at the same linear
stretch employed in the top panels.  Significant residuals can be seen
in either case ({\it right hand panels}) implying that the stream and
spheroid CMDs are not drawn from the same population.}
\label{strmhalomod}
\end{figure}

Given that the stream and spheroid are so similar, we also explored to
what extent both populations might be consistent with a single star
formation history.  First, we simply used the spheroid star formation
history (Figure~\ref{halofit}) to normalize a set of isochrones
scattered according to the stream artificial star tests, and then
scaled the result to match the number of stars in the stream.  This
created a model with the spheroid star formation history but the
observational properties of the stream data, enabling a fair
comparison of the two.  The result is shown in
Figure~\ref{strmhalomod}.  It is obvious that there are gross
residuals in the model.  Although this was not a fit (given that we
simply applied the star formation history of the spheroid), if this
model had resulted from our standard isochrone fitting, it would have
produced a $\chi^2_{\rm eff}$ of 1.32.  The comparison of the spheroid
data with this model population yielded a $\chi^2_{\rm eff}$ of 1.11
(\S\ref{secsph}); the much larger discrepancy of the stream data
with this model population strongly implies that the spheroid and
stream data were drawn from distinct populations, at a confidence
level exceeding 99\%.

\begin{figure}[h]
\epsscale{1.2}
\plotone{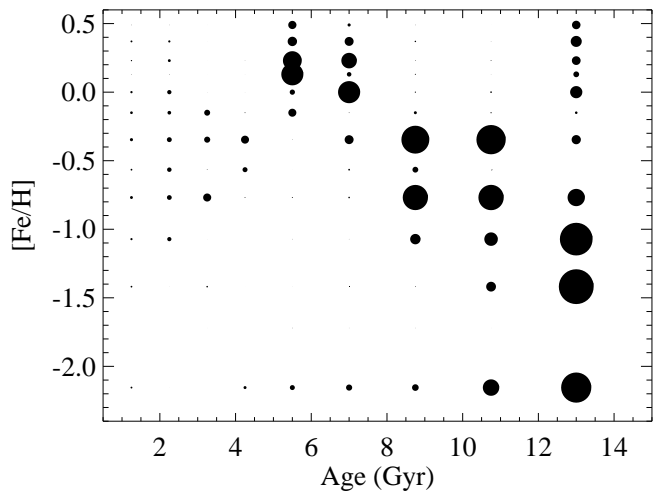}
\epsscale{1.0}
\vskip -0.1in
\caption{
The distribution of age and metallicity in the best-fit model
simultaneously fit to the spheroid and stream data.  The area of the
filled circles is proportional to the number of stars in each
isochrone group.  The distribution shown here is clearly a compromise
between those shown in Figures~\ref{halofit} and \ref{strmfit}.
}
\label{strmhalofit}
\end{figure}

Next, we tried fitting both the spheroid and stream simultaneously
with the same star formation history.  Specifically, a model for the
stream was constructed from isochrones appropriately matching the
stream observations (utilizing the stream artificial star tests), and
a model for the spheroid was constructed from isochrones appropriately
matching the spheroid observations (utilizing the spheroid artificial
star tests), but the relative weights of the isochrones used to
construct these stream and spheroid models came from a single
distribution of age and metallicity.  This distribution was varied
until the best fit to both the stream and spheroid data was achieved.
The resulting age and metallicity distribution is shown in
Figure~\ref{strmhalofit}.  Curiously, this compromise solution to both
CMDs is a bit older and more metal-poor than that found for either CMD
individually; this is likely due to the fact that the spheroid and
stream are distinct, resulting in a poor fit when fitting both at the
same time.  The poor quality of the fit can be seen when this
compromise model is compared to the stream data, as shown in
Figure~\ref{strmhalomod}.  The value for $\chi^2_{\rm eff}$ is not
terrible (1.14), but there are approximately twice the number of
degrees of freedom in this fit, given that we are fitting two CMDs of
data simultaneously, so the deviation from unity is more significant.
Both of these tests imply that while the stream and spheroid CMDs are very
similar, they are not drawn from exactly the same population.

As done with the spheroid data, we also explored to what extent stream models
with more restricted age ranges are consistent with the data.  When
the isochrones are restricted to ages$< 10$~Gyr, the quality of the
fit is nearly unchanged (with $\chi^2_{\rm eff}=1.10$), although the
resulting distribution of age and metallicity looks somewhat skewed,
with much of the weight falling at the top end of the age range.  When
the isochrones are restricted to ages$\ge 10$~Gyr, the quality of the
fit is very poor, with $\chi^2_{\rm eff}=2.80$.  If old metal-rich
stars are removed from the input isochrone grid, the quality of the
fit is unchanged from the best fit model, while if young metal-poor
stars are removed, the quality of the fit is noticeably affected, with
$\chi^2_{\rm eff} = 1.14$, but the model is only
missing the brightest and bluest stars in the blue plume.

\begin{table*}[t]
\begin{center}
\caption{Summary of Stream Fitting}
\begin{tabular}{lcccl}
\tableline
Model & $<$[Fe/H]$>$ & $<$age$>$ & $\chi^2_{\rm eff}$ & Comment \\
\tableline
Standard model & $-0.7$ & 8.8 & 1.08 & Minimal residuals in fit\\
Age $< 10$~Gyr & $-0.6$ & 8.1 & 1.10 & Minimal residuals in fit\\
Age $\ge 10$~Gyr & $-1.0$ & 11.0 & 2.80 & Gross residuals in fit\\
No old metal-rich stars & $-0.7$ & 8.8 & 1.09 & Minimal residuals in fit\\
No young metal-poor stars & $-0.7$ & 8.8 & 1.14 & Misses part of plume\\
Best-fit spheroid model & $-0.6$ & 9.7 & 1.32\tablenotemark{a} & Gross residuals \\
Simultaneous fit to spheroid  stream & $-0.8$ & 10.1 & 1.14\tablenotemark{b} & Significant residuals\\
Fixed 25\% spheroid contamination & $-0.8$ & 8.8 & 1.10 & Similar to standard model\\
\tableline
\multicolumn{5}{l}{$^a$Not actually a fit.  See text for details.}\\
\multicolumn{5}{l}{$^b$Twice the degrees of freedom.  See text for details.}
\end{tabular}
\label{tabstream}
\end{center}
\end{table*}

\begin{figure}[h]
\epsscale{1.2}
\plotone{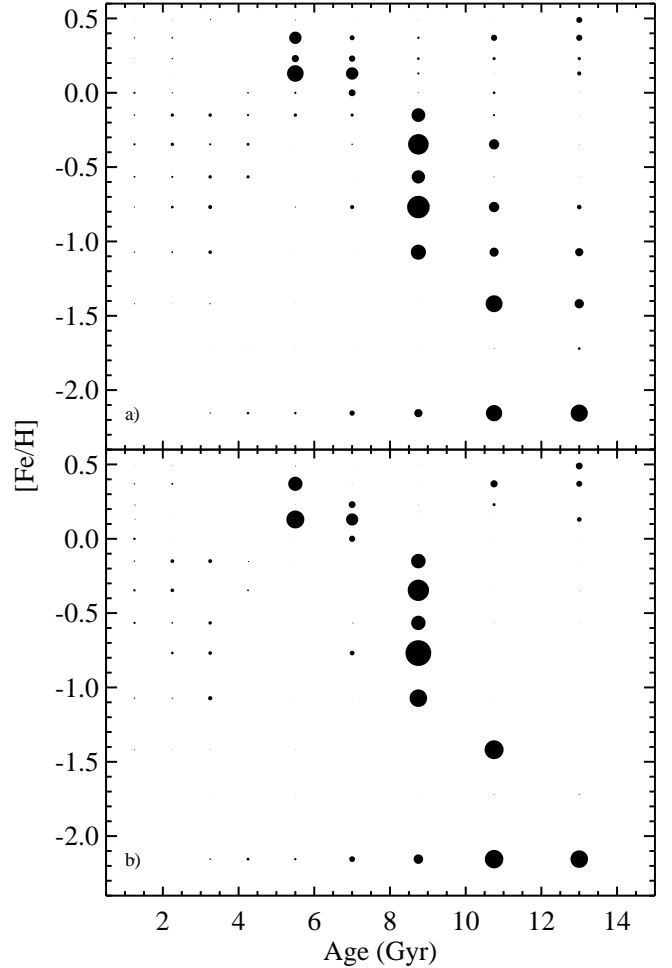}
\epsscale{1.0}
\vskip -0.1in
\caption{ The best-fit model to the stream, assuming a fixed 25\%
contamination from the underlying spheroid that matches the population
in Figure~\ref{halofit}.  The area of the filled circles is
proportional to the number of stars in each isochrone group.  {\it Top
panel:} The complete star formation history, including the fixed
spheroid contamination.  {\it Bottom panel:} The star formation
history for the stream population in isolation, excluding that part of
the fit representing the spheroid contamination.  The population has
been normalized such that the total area in the symbols is the same in
both panels.}
\label{strmfit_cont}
\end{figure}

The Keck data for our stream field imply that 75\% of its stars fall
in two kinematically cold stream components (Kalirai et al.\ 2006b),
and that 25\% of its stars are in the underlying spheroid.  Although
the population in our spheroid field might not be representative of
the underlying spheroid in the stream field, it is reasonable to
wonder how the fitting of the stream star formation history is
affected if this spheroid contamination is taken into account.  To
explore this, we fitted the stream with the same set of isochrones, but
added an additional component to the model, fixed at 25\% of the
population, representing the spheroid contamination.  This
contamination component was constructed from the best-fit model to the
spheroid but using the isochrones scattered with the stream artificial
star tests; thus the contamination component appropriately represents
the spheroid population as it would appear in the stream data.  The
results are shown in Figure~\ref{strmfit_cont}.  The quality of the
fit is good; $\chi^2_{\rm eff} = 1.10$.  In the top panel, we show the
total star formation history (combining the fixed spheroid
contamination and the fit to the stream).  In the bottom panel, we
have subtracted the spheroid contamination component from the star
formation history, to show the star formation history of the stream in
isolation.  The isolated star formation history of the stream
(Figure~\ref{strmfit_cont}b) is very similar to the best-fit model to
the stream that did not try to account for the spheroid contamination
(Figure~\ref{strmfit}).  Given the similarity between the stream and
spheroid CMDs, and the fact that the spheroid contamination is only
25\%, this is not that surprising.  We summarize the results of the
stream fitting in Table~\ref{tabstream}.

\subsection{Results for the Disk}

\begin{figure}[h]
\epsscale{1.2}
\plotone{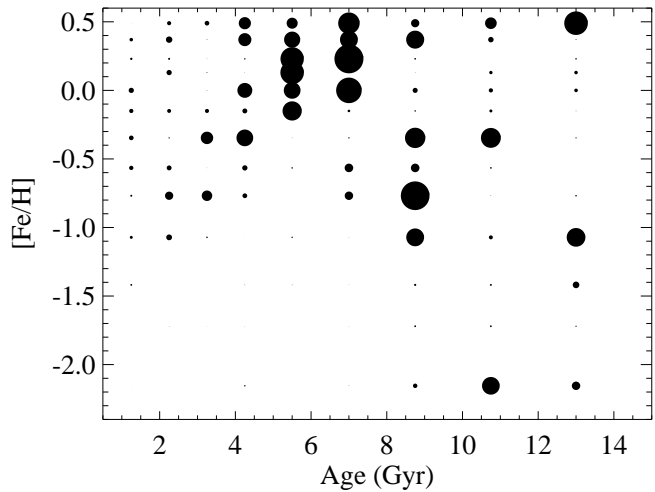}
\epsscale{1.0}
\vskip -0.1in
\caption{The distribution of age and metallicity in the best-fit model
of the disk data.  The area of the filled circles is proportional to
the number of stars in that isochrone group.  The distribution shown
here is clearly distinct from those shown in Figures~\ref{halofit} and
\ref{strmfit}.  The total area within the filled symbols has been
normalized to that in Figures~\ref{halofit} and \ref{strmfit}, to ease
comparisons.}
\label{diskfit}
\end{figure}

\begin{figure}[h]
\epsscale{1.2}
\plotone{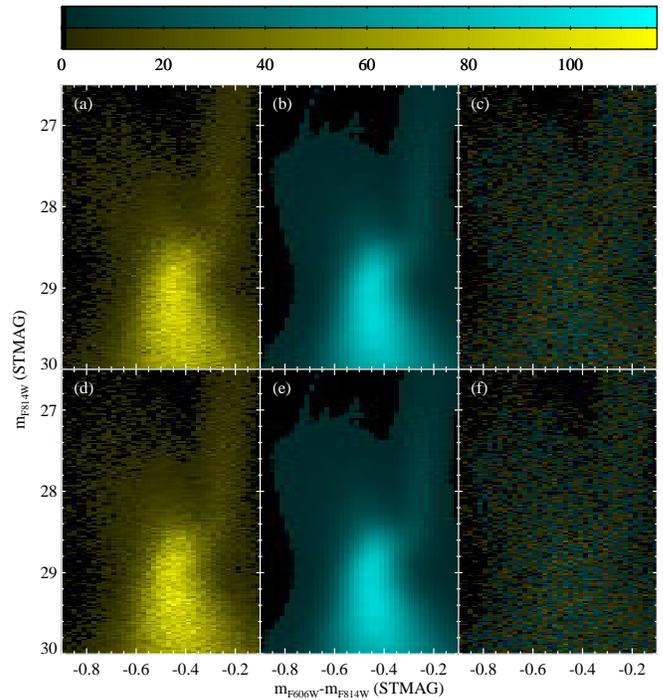}
\epsscale{1.0}
\vskip -0.1in
\caption{{\it Top panels:} The CMD of the disk data ({\it
yellow}), the best-fit model to those data ({\it blue}), and the
differences between the data and model ({\it yellow} and {\it blue}),
all shown at the same linear stretch.  {\it Bottom panels:} An
artificial CMD drawn from the best-fit model ({\it yellow}), the same
best-fit model ({\it blue}), and the differences between the data and
model ({\it yellow} and {\it blue}), all shown at the same linear
stretch employed in the top panels.  Note that the magnitude range
of the disk fit is smaller than that in the spheroid fit, because
the spheroid data are $\sim$0.5 mag deeper than the disk data.}
\label{diskmod}
\end{figure}

The distribution of age and metallicity in our best fit to the disk
data is shown in Figure~\ref{diskfit}.  As expected from our earlier
inspection of the CMDs, the star formation history in the disk is
markedly distinct from that in the spheroid or stream, in the sense
that the population is younger and significantly more metal-rich, with
a mean age of 7.5~Gyr and a mean metallicity of [Fe/H]~=~$-0.2$.  The
fit quality for the best-fit disk model is excellent, with
$\chi^2_{\rm eff} = 1.05$.  In Figure~\ref{diskmod}, we show the
comparison of the best-fit model to the data, as well as the residuals.

The best-fit model to the disk is dominated by stars at ages of less
than 10~Gyr.  In fact, if we fit the data with a subset of the
isochrones restricted to ages$<$10~Gyr (i.e., remove all old
isochrones from the input grid, not just the metal-rich ones), the fit
is only negligibly worse than that achieved with the full set of
isochrones ($\chi^2_{\rm eff} = 1.06$), and the resulting distribution
of age and metallicity looks very similar to that in the best-fit
model.  In contrast, a fit restricted to ages$\ge 10$~Gyr is grossly
inadequate, with $\chi^2_{\rm eff}=5.07$.  If young metal-poor stars
are removed from the fit (as done with the stream and spheroid
fitting), the fit quality drops, with $\chi^2_{\rm eff} = 1.14$, and
the model misses the bright blue stars in the plume.

\begin{table*}[t]
\begin{center}
\caption{Summary of Disk Fitting}
\begin{tabular}{lcccl}
\tableline
Model & $<$[Fe/H]$>$ & $<$age$>$ & $\chi^2_{\rm eff}$ & Comment \\
\tableline
Standard model & $-0.2$ & 7.5 & 1.05 & Minimal residuals in fit\\
Age $< 10$~Gyr & $-0.1$ & 6.9 & 1.06 & Minimal residuals in fit\\
Age $\ge 10$~Gyr & $-0.9$ & 11.0 & 5.07 & Gross residuals in fit\\
No young metal-poor stars & $-0.2$ & 7.6 & 1.14 & Misses part of plume\\
Fixed 33\% spheroid contamination & $+0.1$ & 6.6 & 1.05 & Younger. Minimal residuals\\
\tableline
\end{tabular}
\label{tabdisk}
\end{center}
\end{table*}

The metallicity distribution in our best fit to the disk CMD is
somewhat more metal-rich than that typically found in the outer disk
of M31 (e.g., Worthey et al.\ 2005).  There are several possible
reasons for this.  First, the greatest color dependence upon [Fe/H] is
at the tip of the RGB, which in our data is both sparsely populated
and seriously contaminated by foreground dwarf stars.  Instead, we are
using the lower RGB, which offers the advantage of large numbers of
M31 stars and little contamination, but the penalty is a reduced color
sensitivity to [Fe/H].  Second, the use of distinct isochrone sets and
distinct observing bands results in significant scatter for abundance
determinations even when the population is a simple one, such as a
globular cluster.  The metallicities we derive are calibrated to the
globular cluster metallicities given in Table~\ref{tabclus}.
Published abundances for globular clusters of intermediate metallicity
vary by $\sim$0.2 dex in the recent literature, while abundances for
high metallicity clusters vary by even more (see Brown et al. 2005 and
references therein).  Moreover, isochrones at high metallicity are
difficult to calibrate, given that appropriate clusters tend to be in
heavily reddened regions, such as the Galactic bulge.  Finally,
previous [Fe/H] distributions for M31 fields invariably employed old
isochrones or the ridge lines of old globular clusters as reference
points.  This will bias the results toward lower metallicity if the
metal-rich population is in fact significantly younger than Galactic
globular clusters.  For example, the upper RGB for a 13 Gyr population at
[Fe/H]~=~0.0 is very similar to that for a 6 Gyr population at
[Fe/H]~=~+0.230.

\begin{figure}[h]
\epsscale{1.2}
\plotone{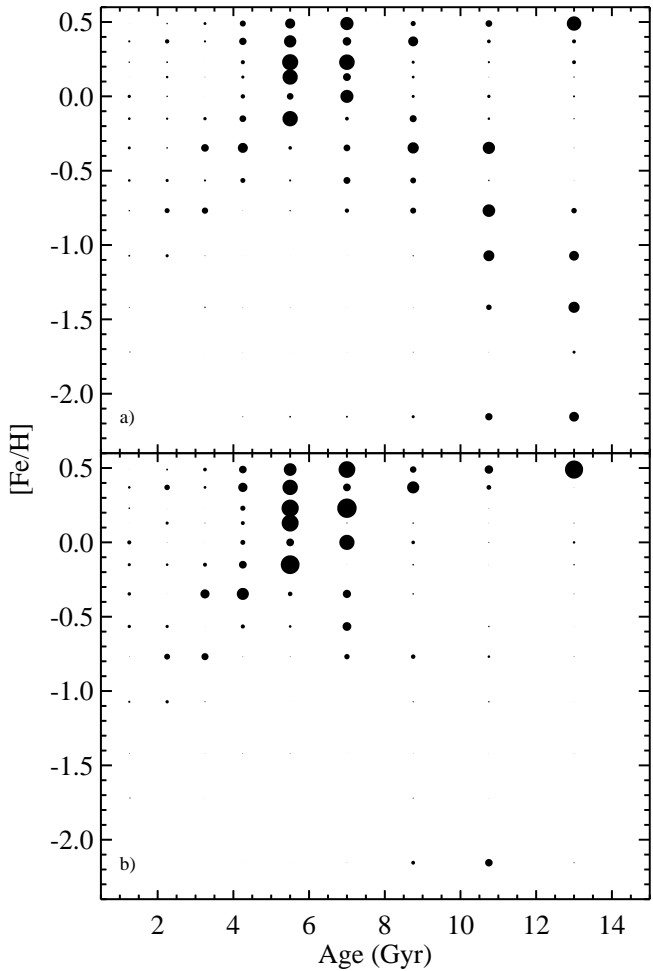}
\epsscale{1.0}
\vskip -0.1in
\caption{ The best-fit model to the disk, assuming a fixed 33\%
contamination from the underlying spheroid that matches the population
in Figure~\ref{halofit}.  {\it Top panel:} The complete star formation
history, including the fixed spheroid contamination.  {\it Bottom
panel:} The star formation history for the disk population in
isolation, excluding that part of the fit representing the spheroid
contamination.  The population has been normalized such that the total
area in the symbols is the same in both panels.}
\label{diskfit_cont}
\end{figure}

The Keck kinematics of our disk field imply that $\sim$67\% of its
stars are moving in the disk (Kalirai et al.\ 2006b; Reitzel et al.\ in
prep.), and that $\sim$33\% of its stars are in the underlying
spheroid.  As with our analysis of the stream, the population in our
spheroid field might not be representative of the underlying spheroid
in the disk field, but it is reasonable to explore a fit to the disk
with a fixed contamination component from the spheroid.  We repeated
the disk fitting with an additional model component held fixed at 33\%
of the population, representing spheroid contamination.  This
contamination component was constructed from the best-fit model to the
spheroid but using the isochrones scattered with the disk artificial
star tests; thus the contamination component appropriately represents
the spheroid population as it would appear in the disk data.  The
results are shown in Figure~\ref{diskfit_cont}.  The quality of the
fit is excellent, with $\chi^2_{\rm eff} = 1.05$. In the top panel, we
show the total star formation history (which includes the fixed
spheroid component in the fit to the disk field).  In the bottom
panel, we show the same fit to the star formation history, but
subtract that fixed component representing spheroid contamination, in
order to show the star formation history of the disk population in
isolation.  The isolated disk population (Figure~\ref{diskfit_cont}b)
is significantly younger and more metal-rich than that found in our
initial model (Figure~\ref{diskfit}), where we did not try to account
for the spheroid contamination.  The isolated point at 13 Gyr and
[Fe/H]~=~0.5 is not significant (repeating the fit with no isochrones
older than 10~Gyr yields $\chi^2_{\rm eff} = 1.06$).  The similarities
between Figures~\ref{diskfit} and Figure~\ref{diskfit_cont}a are
reassuring; the fit in Figure~\ref{diskfit} did not employ any
knowledge of the spheroid contamination, yet it is clear that this fit
tried to reproduce the old metal-poor component that is present in
Figure~\ref{diskfit_cont}a, where we explicitly specified a spheroid
contamination component to the model.  Because the spheroid
contamination can completely account for the old and metal-poor stars
in the disk field, the dearth of metal-poor stars is another example
of the ``G dwarf problem'' -- that a simple closed box model of
chemical evolution predicts a longer tail of metal-poor stars than
seen in all massive galaxies (see Worthey et al.\ 2005 and references
therein). We summarize the results of the disk fitting in
Table~\ref{tabdisk}.

\subsection{Systematic Effects of Binaries, Alpha-enhancement, 
IMF, Distance, and Reddening}
\label{syssec}

The fits above make assumptions about the binary fraction,
alpha-element enhancement, IMF, distance, and reddening.  Of these
three parameters, the binary fraction and reddening uncertainties
translate into the largest uncertainties in the resulting fits, but do
not change the gross interpretation of the CMDs.  We assumed an IMF
index of $-1.35$ (Salpeter 1955), and assumed that [$\alpha$/Fe]~=~0.3
at [Fe/H]~$\leq -0.7$ and [$\alpha$/Fe]~=~0.0 at [Fe/H]~$> -0.7$.  We
assumed 770~kpc for the M31 distance (Freedman \& Madore 1990), which
is based on Cepheids and falls in the middle of the range generally
quoted in the literature (e.g., Pritchet \& van den Bergh 1987; Stanek
\& Garnavic 1998; Holland 1998; Durrell et al.\ 2001; McConnachie et
al.\ 2005; Ribas et al.\ 2005).  
We assumed $E(B-V)=0.08$~mag in each field, but as noted
earlier, the Schlegel et al.\ (1998) map is uncertain at the
$\sim$0.02~mag level in random fields, with somewhat higher
uncertainties near Local Group galaxies.

We chose a binary fraction of 10\%, because grossly changing this
value produced lower quality fits with obvious residuals in the
comparison of the models and data.  Given that we chose a binary
fraction that minimized fit residuals, in a sense we ``fit'' the
binary fraction, but did so on a very coarse scale.  Fortunately all
three fields can be reasonably fit with the same binary fraction,
because this avoids complications in the interpretation of the fits.
If we assumed distinct binary fractions in the fitting to each field,
one could attribute some of the age variations to this varying binary
fraction.  At larger binary fractions, the features in the synthetic
CMD become brighter, and the age distribution must shift to older ages
to compensate, while lower binary fractions result in younger age
distributions.

\begin{figure*}[t]
\epsscale{1.1}
\plotone{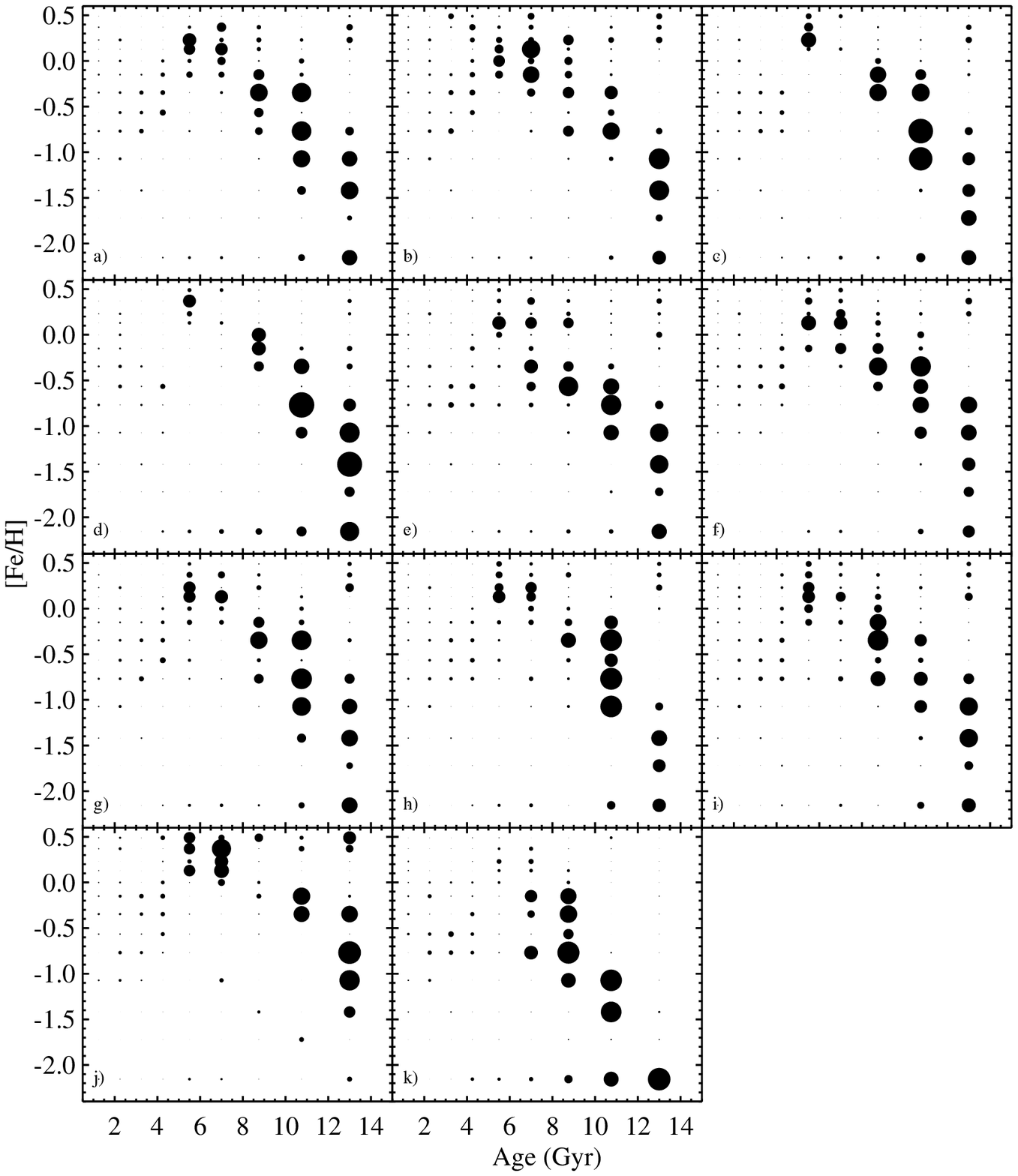}
\epsscale{1.0}
\caption{The distribution of age and metallicity in the best-fit model
to the spheroid data, making different assumptions about the binary
fraction, IMF, alpha-enhancement, distance, and reddening.  The area
of the filled circles is proportional to the number of stars in each
isochrone group.  {\it a)} Our standard model: binary fraction 10\%,
Salpeter (1955) IMF, alpha-element enhancement at [Fe/H]$<-0.7$,
distance of 770 kpc, $E(B-V)=0.08$~mag.  {\it b)} A binary fraction
of 0\%.  {\it c)} A binary fraction of 20\%.  {\it d)} A binary
fraction of 40\%.  {\it e)} Isochrones at [Fe/H]$\leq 0$ are
alpha-enhanced.  {\it f)} None of the isochrones are alpha-enhanced.
{\it g)} An IMF index of $-1.15$.  {\it h)} Distance is 760~kpc.  {\it
i)} Distance is 780~kpc.  {\it j)} $E(B-V)=0.05$~mag.  {\it k)}
$E(B-V)=0.11$~mag.  }
\label{sysfig}
\end{figure*}

To demonstrate the sensitivity of our fits to these parameters, we
repeated our fits while varying our assumptions.  The results are
shown in Table~\ref{systab} for all three fields, and in
Figure~\ref{sysfig} for the spheroid field.  Reducing the binary
fraction to 0\% would decrease our ages by 0--0.4~Gyr, while
increasing the binary fraction to 40\% would increase our ages by
$\sim$1~Gyr.  Changing the alpha enhancement has almost no effect,
other than a slight shift in the metallicity distribution.  The
insensitivity to alpha enhancement makes sense, because in these
bandpasses, isochrones with enhanced alpha elements look much like
scaled-solar isochrones at slightly higher metallicity (note that the
isochrones are always transformed to the ACS bandpasses using
synthetic spectra of a consistent alpha enhancement; see Brown et al.\
2005).  Changing the IMF index from $-1.35$ to $-1.15$ also has little
effect on the metallicity and age distributions; this is
because our CMDs are sampling a fairly small range in stellar mass
(the bulk of the stars brighter than the faint limit in our fitting
region fall in the mass range 0.7$\lesssim M \lesssim 1.2$~$M_\odot$).
Changing the extinction by 0.03~mag in either direction (assuming the
average Galactic extinction curve of Fitzpatrick 1999) primarily
affects the metallicity distribution; an increase in the assumed
extinction (redder stars) is compensated by a lower metallicity (bluer
stars), and vice versa.  Changing the distance modulus by 0.03~mag in
either direction primarily affects the age distribution; an increase
in the assumed distance (fainter apparent magnitudes) is compensated by
a younger age (brighter absolute and apparent magnitudes), and vice
versa.  Note that no change in assumptions for the spheroid
(Figure~\ref{sysfig}) can make the spheroid population look like that
of the disk (Figure~\ref{diskfit}).

\begin{table*}[t]
\begin{center}
\caption{Systematic Effects of Assumptions}
\begin{tabular}{lccccccccc}
\tableline
    &  \multicolumn{3}{c}{spheroid} &  \multicolumn{3}{c}{stream} &  \multicolumn{3}{c}{disk} \\
Fit & $<$age$>$ & $<$[Fe/H]$>$ & $\chi^2_{\rm eff}$ & $<$age$>$ & $<$[Fe/H]$>$ & $\chi^2_{\rm eff}$ & $<$age$>$ & $<$[Fe/H]$>$ & $\chi^2_{\rm eff}$ \\
\tableline
Standard model                         &  9.7 & $-$0.6 & 1.11 & 8.8 & $-$0.7 & 1.08 & 7.5 & $-$0.2 & 1.05 \\
Binary fraction = 0.0                  &  9.3 & $-$0.5 & 1.14 & 8.7 & $-$0.7 & 1.09 & 7.5 & $-$0.1 & 1.11 \\
Binary fraction = 0.2                  & 10.1 & $-$0.8 & 1.17 & 9.2 & $-$0.8 & 1.13 & 7.7 & $-$0.2 & 1.05 \\
Binary fraction = 0.4                  & 10.9 & $-$0.9 & 1.42 & 9.9 & $-$1.0 & 1.25 & 8.5 & $-$0.4 & 1.13 \\
$[\alpha$/Fe$]=0.3$ at [Fe/H]$\leq$0   &  9.7 & $-$0.7 & 1.11 & 8.9 & $-$0.8 & 1.08 & 7.7 & $-$0.3 & 1.05 \\
$[\alpha$/Fe$]=0.0$ at all [Fe/H]      &  9.7 & $-$0.6 & 1.11 & 8.9 & $-$0.7 & 1.08 & 7.6 & $-$0.1 & 1.05 \\
IMF index $-$1.15                      &  9.9 & $-$0.7 & 1.11 & 9.0 & $-$0.7 & 1.08 & 7.6 & $-$0.2 & 1.04 \\
Distance = 760 kpc                     &  9.8 & $-$0.7 & 1.12 & 9.1 & $-$0.7 & 1.09 & 7.6 & $-$0.2 & 1.04 \\ 
Distance = 780 kpc                     &  9.5 & $-$0.6 & 1.11 & 8.7 & $-$0.7 & 1.08 & 7.4 & $-$0.1 & 1.05 \\
$E(B-V)=0.05$~mag                      &  9.8 & $-$0.2 & 1.15 & 9.1 & $-$0.3 & 1.08 & 7.7 & $+$0.2 & 1.14 \\
$E(B-V)=0.11$~mag                      &  9.4 & $-$1.0 & 1.17 & 8.6 & $-$1.1 & 1.16 & 7.3 & $-$0.5 & 1.05 \\
\tableline
\end{tabular}
\label{systab}
\end{center}
\end{table*}

\section{Discussion}
\label{secdisc}

\begin{figure}[h]
\epsscale{1.1}
\plotone{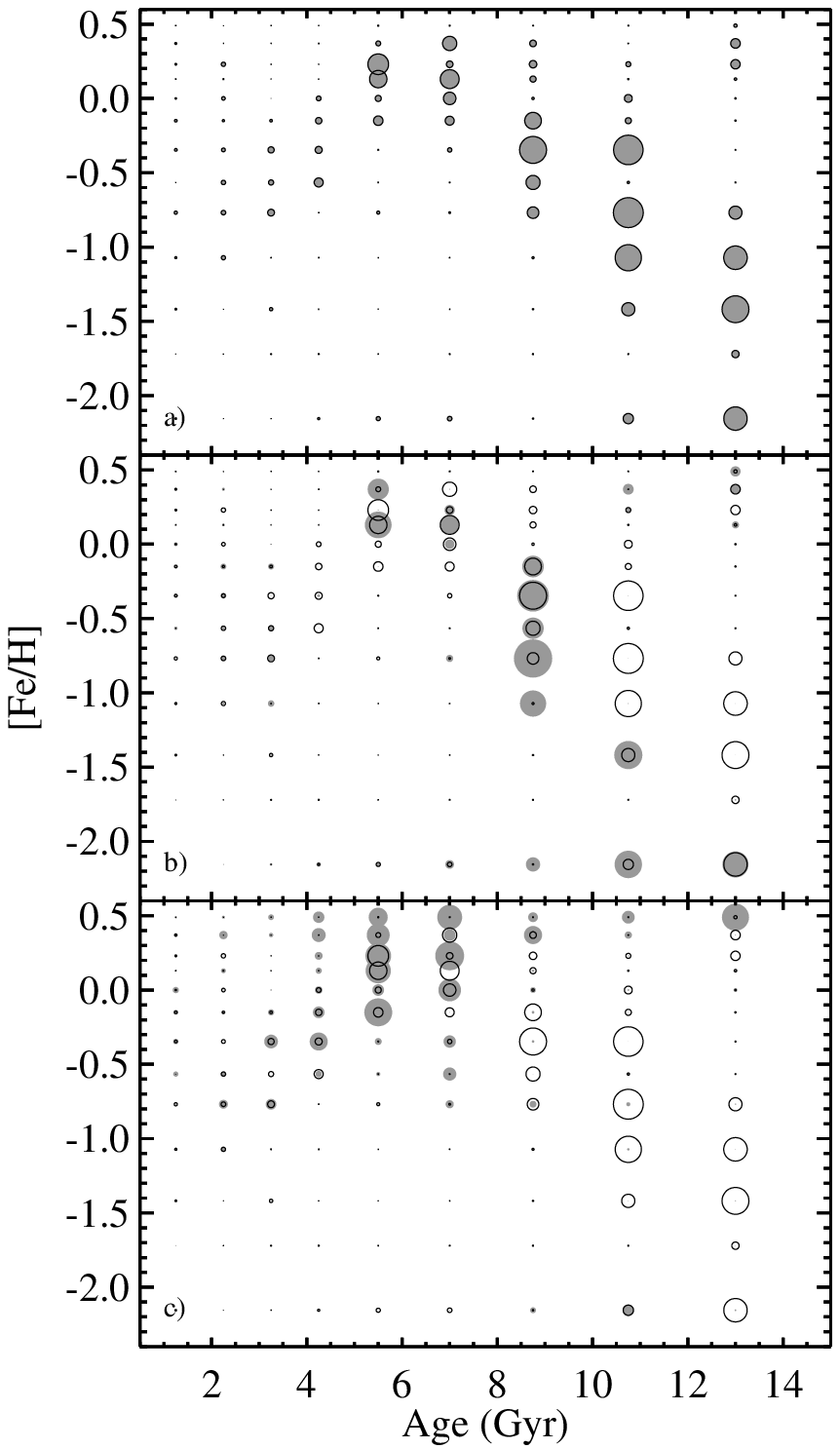}
\epsscale{1.0}
\caption{The best-fit star formation histories for the
spheroid ($a$), stream ($b$), and disk ($c$).  The area
of the filled circles ({\it grey}) is proportional to the
number of stars falling in the given isochrone.  For comparison,
the star formation history of the spheroid is overplotted in
each panel ({\it black open circles})  The stream and disk fits
each assumed a fixed contamination from the spheroid, which has
been subtracted.
}
\label{compfits}
\end{figure}

The quantitative fitting to the CMDs of the spheroid, stream, and
outer disk reaffirmed our general impressions from the qualitative
inspection of the CMDs.  In Figure~\ref{compfits}, we compare the star
formation histories for the three fields.  The star formation history
in the spheroid is simply our standard model (Figure~\ref{halofit}),
while the star formation histories in the stream and disk are those
that have had an assumed spheroid contamination subtracted
(Figures~\ref{strmfit_cont}b and \ref{diskfit_cont}b).  All three
fields show an extended star formation history.  The star formation
history in the stream is similar to that in the spheroid, but is
shifted somewhat younger.  The disk population is dominated by
intermediate-age stars, with little evidence for the old metal-poor
population present in the spheroid and stream.  All three fields have
a trace population of young metal-poor stars, presumably due to the
accretion of metal-poor stars from dwarf galaxies or due to stars
forming from the infall of relatively pristine material.  The fact
that such material continues to fall into Andromeda is evidenced by the
extensive population of \ion{H}{1} clouds recently found in the
outskirts of the galaxy (Thilker et al.\ 2004).

\subsection{Disk}

Most hierarchical CDM models predict that a spiral disk forms
inside-out, generally leading to a disk that becomes progressively
younger at increasing radius.  For example, the simulated disk of
Abadi et al.\ (2003a, 2003b) has a mean age of $\sim 8$--10~Gyr within
2 kpc of the center and $\sim$6--8~Gyr near 20 kpc.  However, the
literature does include counter-examples with more complex age
gradients.  The simulated galaxy of Robertson et al.\ (2004) exhibits
a mean stellar age of $\sim$7.5~Gyr in the center (within 2 kpc) and
$\sim$10~Gyr in the disk outskirts (beyond 14 kpc).  The CDM models of
Sommer-Larson, G$\ddot{\rm o}$tz, \& Portinari (2003) result in disk
galaxies that sometimes form inside-out and sometimes form outside-in.
Both classes predict mean ages of 6--8~Gyr in the outer disk (6 scale
lengths from the center), but the age distributions differ, with the
inside-out galaxy hosting a significantly larger fraction of young
stars ($\lesssim 3$~Gyr) in the outskirts.  In a sophisticated model
of the chemical evolution in the Milky Way disk, Chiappini et al.\
(2001) demonstrate an inside-out formation scenario where the stellar
age is not a monotonically varying function of distance from the
Galactic center; in the inner disk (4--10 kpc), the stellar ages are
decreasing with increasing radius, as expected, but beyond this radius,
the stellar ages increase with radius, because the thick disk and halo
begin to dominate over the thin disk.  All of these models can be
compared to the solar neighborhood (e.g., Ibukiyama \& Arimoto 2002;
Sandage, Lubin, \& VandenBerg 2003; Fontaine, Brassard, \& Bergeron
2001), but we know little of the detailed star formation histories for
other giant spiral galaxies.  As far as the structures are concerned,
observations of high-redshift disk galaxies (e.g. Ferguson, Dickinson
\& Williams 2000; Ravindranath et al.\ 2004) suggest that disks were
largely in place 8 Gyr ago.  Since then, they have increased their
stellar masses and increased their sizes consistent with an inside-out
sequence of star formation (Trujillo et al.\ 2005), with the average
stellar surface mass density staying roughly constant from $z=1$ to
the present (Barden et al.\ 2005).

Our mean age in the outer disk (6.6~Gyr) is in good agreement with the
models of Abadi et al.\ (2003a, 2003b), and significantly younger than
the models of Robertson et al.\ (2004); these comparisons
suggest a consistency with an inside-out formation scenario.  Our mean
age also falls in the range found in both the inside-out and
outside-in models of Sommer-Larson et al.\ (2003), but our age
distribution, with a significant dearth of stars younger than 3~Gyr,
is in somewhat better agreement with their outside-in model.  However,
these are all hydrodynamical models that track the birth of
particles but largely ignore the details of chemical evolution. It
would be interesting to compare our age-metallicity distribution with
such a distribution in a true chemical evolution model under an
inside-out formation scenario (e.g., Chiappini et al.\ 2001).  

\begin{figure}[ht]
\epsscale{1.1}
\plotone{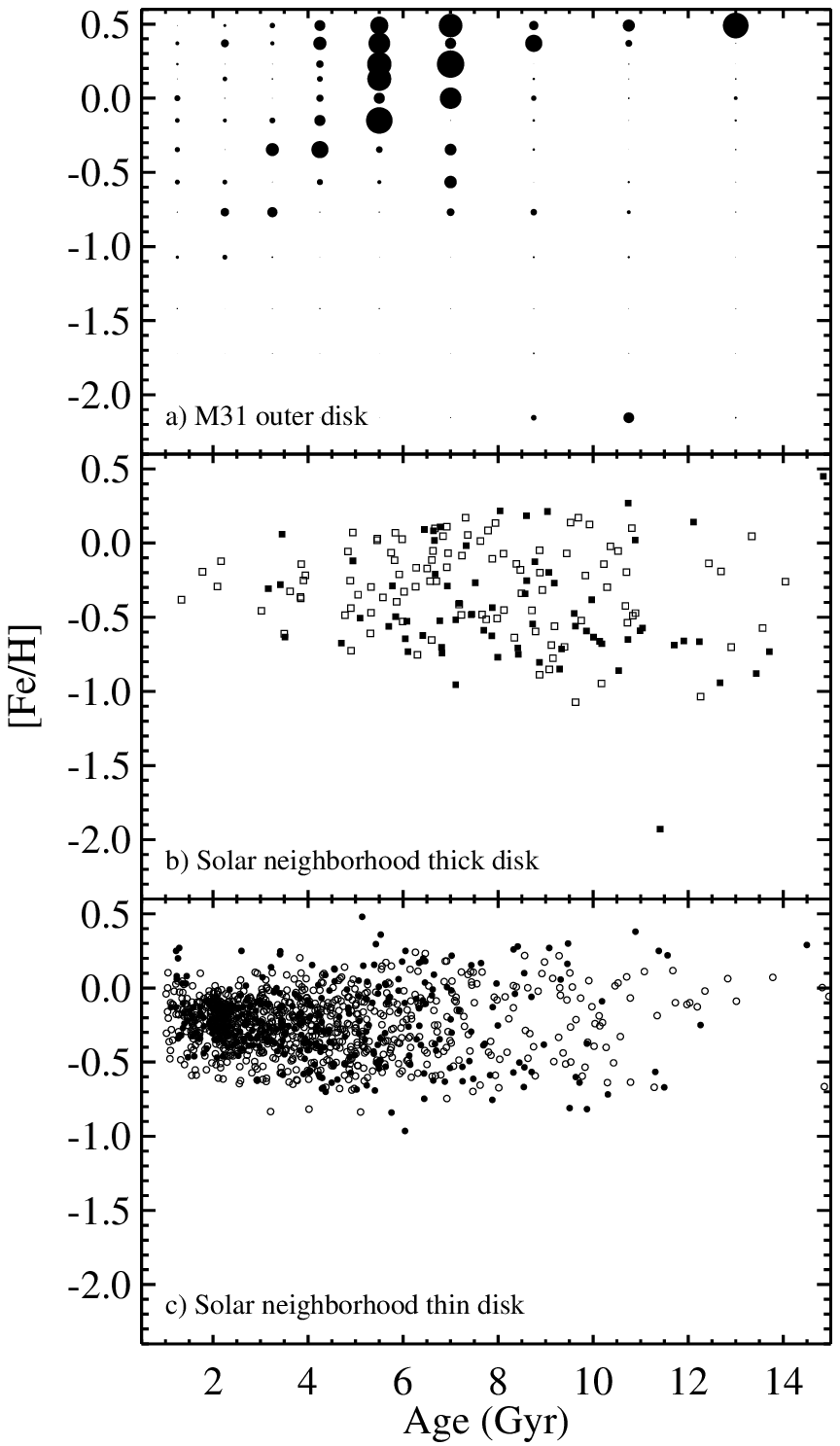}
\epsscale{1.0}
\caption{{\it a)} The distribution of age and metallicity in the
best-fit model of the disk data (assuming a 33\% contamination from
the spheroid, which has been subtracted).  The area of the filled circles is
proportional to the number of stars in that isochrone group.  The
distribution shown here is clearly distinct from those shown in
Figures~\ref{halofit} and \ref{strmfit}.  {\it b)} The distribution of
age and metallicity for individual thick disk stars in the solar
neighborhood, from the photometric ({\it open boxes}) and
spectroscopic ({\it filled boxes}) measurements of Ibukiyama \&
Arimoto (2002; their Figure~8).  {\it b)} The distribution of age and
metallicity for individual thin disk stars in the solar neighborhood,
from the photometric ({\it open circles}) and spectroscopic ({\it
filled circles}) measurements of Ibukiyama \& Arimoto (2002; their
Figure~5).}
\label{diskfit_comp}
\end{figure}

Our star formation history in the disk is probably saying less about
the validity of the inside-out formation scenario and more about the
relative scales of the thin and thick disk; because our disk field is
25~kpc from the galactic center, it is well into the regime where one
might expect the thick disk to dominate (Chiappini et al.\ 2001).
Indeed, there is evidence that the thick disk begins to dominate well
inside this radius; in their WFPC2 images of an off-axis field 5~kpc
from the nucleus, Sarajedini \& Van Duyne (2001) found a population
apparently dominated by thick disk stars.
Note that Morrison et al.\ (2004) apparently found a subsystem of the
M31 globular cluster system with thin disk kinematics, but this
subsystem is largely restricted to that part of the disk plane
interior to our own disk field.  In Figure~\ref{diskfit_comp}, we
compare the age and metallicity distribution in our disk field to
those distributions in the solar neighborhood.  The outer disk of
Andromeda is clearly similar to the thick disk population of the solar
neighborhood (dominated by intermediate-age stars at relatively high
metallicities), but looks nothing like the thin disk of the solar
neighborhood (dominated by stars younger than 5~Gyr).  The hydrogen
column density in our disk field (Table~\ref{fieldtab}; Braun et al.\
in prep.) is below the threshold typically assumed for star formation
in disk galaxies ($N_{HI} \sim 10^{21}$ cm$^{-2}$; Kennicutt 1989),
and so the dearth of very young stars should not be surprising.

Our star formation history in the disk is in rough agreement with that
found by other groups studying the outskirts of the disk with
shallower {\it HST} data.  Looking at a field $\sim$15$^{\prime}$
further away from the galaxy center than our own field, Ferguson \&
Johnson (2001) found a somewhat older and more metal-poor population;
they quoted a mean age $\gtrsim 8$~Gyr and a metallicity of
$<$[Fe/H]$>$~$\sim -0.7$.  They reported trace populations of young
stars ($\sim 1.5$--3 $M_\odot$) and ancient metal-poor stars ($\gtrsim
10$~Gyr and [Fe/H]$\sim -1.7$), which we also find in our field.
Ferguson \& Johnson (2001) assumed that disk stars comprised
$\sim$95\% of their field population, based on an extrapolation of
the Walterbos \& Kennicutt (1988) decomposition.  During our
observation planning, we also used the work of Walterbos \& Kennicutt
(1988) as a guide, and estimated that the disk contribution in our own
field was similarly high.  We were subsequently surprised to find that
the kinematic data in our field imply that the disk in fact comprises only
67\% of the population (Figure~\ref{cmdsvels}); it must be even lower
in the Ferguson \& Johnson (2001) field.  The disk is clearly falling
off more rapidly than an extrapolation of the Walterbos \& Kennicutt
(1988) data from the interior.  Note that on the other side of the
galaxy, looking in the outer disk near the massive cluster G1, Rich et
al.\ (2004) also found a population dominated by intermediate-age
stars (6--8~Gyr).  The dominance of intermediate-age stars in the
outer disk of Andromeda appears to be ubiquitous.

Looking at fields sampling a wide range of radial distance and
azimuthal angle in Andromeda, Ibata et al.\ (2005) found significant
numbers of stars moving with velocities close to the expected mean
velocity for circular orbits.  They found these stars primarily at
distances of 15--40~kpc from the center, with possible detections out
to 70~kpc.  Their extended disk has an exponential scale length of
5.1~kpc, similar to that of the bright inner disk, but its irregular
morphology and substructure strongly suggest that it is dominated by tidal
debris.  They estimate that the luminosity of this ``disk-like
structure'' accounts for $\sim$10\% of the total luminosity in the M31
disk.  For reference, their ``F13'' field is near our outer disk field
($\sim$10$^{\prime}$ away), and shows kinematic structures very
similar to those in Figure~\ref{cmdsvels}c; their data show a narrow
peak near the velocity expected for stars orbiting in the disk, and a
much broader peak for spheroid stars that show little rotation with
the disk.  Ibata et al.\ (2005) argue that their extended disk is more
likely associated with the thin disk than the thick disk of Andromeda.
However, given the kinematic and population data in our outer disk field, it
would seem more likely that 
their disk-like structure is an extension of the thick
disk.  This would also be consistent with its irregular morphology,
given that thick disks are thought to form via mergers that disrupt
the thin disk (see Wyse et al.\ 2006 and references therein). \\

\subsection{Spheroid and Stream}

As found by Brown et al.\ (2003), the Andromeda spheroid population
spans a surprisingly wide range of age and metallicity, especially
compared to the halo of the Milky Way.  Given the substructure in
Andromeda (Ferguson et al.\ 2002; Figure~\ref{mapfig}) and the success
of $\Lambda$CDM models, we have strong observational and theoretical
reasons for turning to merger scenarios as possible explanations for
the observed distribution of age and metallicity.  One can imagine
that, compared to the Milky Way, Andromeda has experienced many more
small mergers or a few more large ones.  These mergers may have
polluted the inner spheroid with their own material and material
from the Andromeda disk and bulge; in this scenario, the declining
presence of this pollution at increasing radius would account
for the appearance of the spheroid beyond 30~kpc, which looks
more like a canonical metal-poor halo (Guhathakurta et al.\ 2005; 
Irwin et al.\ 2005).

If the Andromeda spheroid is the result of many smaller mergers that
did not occur in the Milky Way, one must ask why there is such a
statistically significant distinction between the merger histories of
two similarly-sized spirals in the same galaxy group.  Is Andromeda
the ``normal'' massive spiral, having cannibalized 10 small galaxies
in its history, while the Milky Way is a 3$\sigma$ outlier, having
cannibalized only 1 small galaxy?  Alternatively, if the Andromeda
spheroid was polluted by one large merger that did not occur in the
Milky Way, one may ask if such a merger is consistent with the
disturbed, but not destroyed, Andromeda disk.  Plausible merger
scenarios must balance both of these concerns.

Recent models by Font et al.\ (2006a) show promise in this regard.  In
their various realizations of a spiral galaxy halo, two models stand
out.  One halo underwent a large accretion event ($10^{8-9}$ $M_\odot$ stellar
mass) 11 Gyr ago, and the other underwent two accretion events ($10^9$
$M_\odot$ stellar mass) $\sim$8.5~Gyr ago; in the former case, the resulting
halo had a lower mean metallicity, with
$<$[Fe/H]$>$=$-1.3$, while in the latter case, the resulting
halo had a significantly higher mean metallicity, with $<$[Fe/H]$>$=$-0.9$.
Vel$\acute{\rm a}$zquez \& White (1999) find that, depending upon the
orbit of the infalling satellite, satellites with up to 20\% of the
disk mass can be accreted without destroying the disk.  Clearly the
amount of disk disruption spans a continuum of outcomes depending upon
the mass of the infalling satellite and its orbit.  Given a mass of
$\approx 7 \times 10^{10}$ $M_\odot$ in Andromeda's disk (Geehan et
al.\ 2006), the disk could survive the accretion of one or two
$\sim$10$^{9-10}$ $M_\odot$ satellites that would in turn
significantly increase the spheroid metallicity.  It is worth noting
that in the Font et al.\ (2006a) models, when metal-rich stars are
present in the spheroid, they are still predominantly old, whereas the
metal-rich stars are very clearly of intermediate age in our own data.
With only 11 of these computationally-intensive realizations, it
appears that the Font et al.\ (2006a) simulations do not sufficiently
populate the possible parameter space to demonstrate if these old
metal-rich stars are a fluke or a general tendency in the models.
In contrast, recent simulations by Renda et al.\ (2005) show that spiral
galaxies with more extended merging histories can have halos that are
both younger and metal-rich.
Could the distinction between the spheroids of the Milky Way and
Andromeda be due to the ingestion of something like the LMC?  There is
also evidence that the globular cluster system of Andromeda includes
clusters much younger than those in our own Galaxy, although it is
debatable if these clusters could have originated in the accretion of
something like the LMC (e.g., Puzia et al.\ 2005; Burstein et al.\
2004; Beasley et al.\ 2005), which hosts a large globular cluster
system spanning a wide range of ages.

Andromeda is not alone in having a metal-rich spheroid with an age
dispersion.  The halo of NGC5128 (Cen A) is metal rich, with
$<$[Fe/H]$> = -0.41$ (Harris, Harris, \& Poole 1999).  The presence of
long period variables with extremely long periods (Rejkuba et al.\
2003) implies the presence of young stars, while the analysis of the
HB, RGB, and AGB populations found in deep {\it HST} photometry of the
galaxy imply an average age of $\sim$8~Gyr in its halo (Rejkuba et
al.\ 2005).  The galaxy also shows evidence for mergers in its shells
and dust lane (Malin, Quinn, \& Graham 1983).

The relatively high metallicity of the stream implies its progenitor
was at least as massive as $10^9$ $M_\odot$ (see Dekel \& Woo 2003);
as such, most numerical simulations of the stream assume it is a dwarf
galaxy that was only recently disrupted by close passage to Andromeda,
within the last $\sim$0.5~Gyr (Font et al.\ 2006b; Fardal et al.\
2006).  The star formation history in the stream is plausible for such
a progenitor, given the wide range of star formation histories seen in
Local Group dwarfs (Mateo 1998).  As noted by Brown et al.\ (2006), it
would be worth exploring whether or not the progenitor is a disk
galaxy, given that the stream combines a relatively high metallicity
with a low velocity dispersion; however, models by Font et al.\
(2006b) and Fardal et al.\ (2006) imply this discrepancy in velocity
can perhaps be explained by dynamical cooling.

The strong similarities between the spheroid and stream populations
offer another clue, but it is a puzzling one.  The field population of
the Milky Way halo does not look to be comprised of populations like
those of present-day dSphs (Shetrone et al.\ 2003), but the field
population of the Andromeda spheroid looks nearly identical to that of
one of its infalling satellites.  A natural question is whether the
$10^{9-10}$ $M_\odot$ merger needed to explain the spheroid data is
sitting in plain sight: the stream.  However, if the progenitor of the
stream really is on its first or second orbit around the galaxy, with
much of its debris coherent on the sky, it is unlikely to comprise a
significant fraction of the population in the relatively smooth
regions of the spheroid, such as our field.  As noted by Brown et al.\
(2006), the star count map of Andromeda (Figure~\ref{mapfig}) and the
kinematic data (Figure~\ref{cmdsvels}) imply that the stream dominates
over the spheroid by a 3:1 ratio in our stream field, but these same
data show no evidence for a single dominant stream in our spheroid
field.  Current orbit models for the stream span a wide range of
possibilities (e.g., Font et al.\ 2006b; Fardal et al.\ 2006); even if
the stream wraps around the Andromeda nucleus and then passes through
our spheroid field (e.g., Ibata et al.\ 2004), it is implausible that
it would spread out enough to hide in the star count maps and
kinematic data, yet still comprise $\sim$75\% of the population in our
spheroid field.  Furthermore, the metallicity distribution in our
spheroid field is clearly very similar to the metallicity distribution
in other fields throughout the inner spheroid of Andromeda (Ferguson
et al.\ 2002; Durrell et al.\ 1994, 2001, 2004).  
Thus, arguments (e.g., Ibata et al.\ 2004) that the intermediate-age 
metal-rich stars in our spheroid field simply represent contamination 
by the stream would seem to imply
that the inner spheroid is metal-rich and ancient everywhere except
for our spheroid field, where the $\sim$40\% of the population is
metal-rich and of intermediate age.  Instead of invoking such a
conspiracy, it is much more plausible that the high metallicities seen
throughout the inner spheroid are associated with intermediate-age
populations, as in our particular spheroid field.

The modeling of the stream's progenitor and its possible orbits is
still in the early stages.  Can a model be constructed where the
debris of the stream progenitor dominates the relatively smooth inner
spheroid everywhere, while maintaining a coherent tidal tail on the
sky?  At the moment, models for the stream progenitor are focused on a
$\sim$10$^9$ $M_\odot$ dwarf galaxy progenitor that only recently
merged with Andromeda (within the last few hundred Myr).  How far can
the models be pushed away from this scenario?  At what point does the
disruption of the Andromeda disk exceed the level of substructure seen
by Ferguson et al.\ (2002)?  Depending upon the orbit, the progenitor
could be as massive as a few $10^{10}$ $M_\odot$ without destroying
the Andromeda disk.  If the progenitor was significantly more massive
than the $10^9$ $M_\odot$ typically assumed now, and perhaps an
infalling disk galaxy, could the start of the merger be pushed
backward in time, such that its debris could more fully pollute the
inner spheroid while still leaving a coherent debris stream on the
sky?  Alternatively, the pollution of the inner spheroid might be due
to a merger event unrelated to that which produced the stream.  The
recent models of Penarrubia, McConnachie, \& Babul (2006) are
interesting in this regard; they find that an ancient merger with a
massive dwarf (10$^{9-10}$ $M_\odot$) could produce the extended
disk-like population found by Ibata et al.\ (2005).

Brown et al.\ (2006) offered two other possible explanations for the
stream and spheroid similarities, but noted that they were
problematic.  One possibility is that the spheroid is comprised of
many disrupted satellites similar to the stream progenitor.  However, it is
difficult to see how the ensemble average of these disrupted
satellites (the spheroid) would so closely resemble the population
in a single disrupted satellite (the stream).  Although the star
formation history for the stream is plausible for a dwarf galaxy,
it is not plausible that it is representative for all dwarf
galaxies already cannibalized by Andromeda.  Another possibility is
that the stream is comprised of material disrupted from the Andromeda
disk and that the same event polluted the spheroid, but it is unclear
if the dynamics and energetics of such a scenario can actually work,
and the stellar populations in our three fields offer evidence against this
scenario (Figure~\ref{compfits}).  The isolated disk population
(removing the spheroid contamination) is dominated by metal-rich
($-0.5 < $ [Fe/H] $< +0.5$) intermediate-age (4--8 Gyr) stars.  The
isolated stream population (removing the spheroid contamination), on
the other hand, also contains stars that are both older and more
metal-poor.  If our disk population is representative of the outer
disk in general, creating the stream from a disruption of disk material
would not result in a stream hosting so many old and metal-poor stars.
This does not preclude significant contamination of the spheroid by
disrupted disk stars -- the population mix in our spheroid field might
be an older metal-poor halo with some contribution of disrupted disk
stars -- but we are still left with coincidence to explain the
similarity between the stream and spheroid populations.

\subsection{Does the Disk Contribute to our Spheroid Field?}

Recently, Worthey et al.\ (2005) put forth a provocative hypothesis,
based on chemical evolution arguments and the high metallicity of the
Andromeda spheroid: that all fields in the spheroid observed to date
are actually dominated by the disk.  They suggested that this
hypothesis could explain the surprisingly broad range of ages found in
our spheroid field (Brown et al.\ 2003).  More recently, Ibata et al.\
(2005) found stars 40 kpc from the center of Andromeda (in all
directions) that appear to be moving in the disk.  With the kinematic
and population information available, we can show that the disk
contribution in our spheroid field must be very small ($\lesssim
1$\%), as originally claimed by Brown et al.\ (2003).

The relevant data are in Figure~\ref{cmdsvels} and
Table~\ref{fieldtab}.  Given the disk inclination of 12.5$^{\rm o}$,
our spheroid field is 11 kpc from the galactic center in the plane of
the sky and 51 kpc from the center in the plane of the disk.  The disk
field is 25 kpc from the galactic center in both the plane of the sky
and the plane of the disk.  

Figure~\ref{cmdsvels}c shows the distribution of velocities in our
disk field.  There are clearly two components.  The broader component
(comprising $\sim$1/3 of the population) is at the systemic velocity
of Andromeda, while the narrower component is redshifted with respect
to Andromeda due to the rotation of the Andromeda disk.  In the
Worthey et al.\ (2005) scenario, one would associate the broad
component with the thick disk and the narrow component with the thin
disk, with only the latter component significantly rotating.  However,
we know from the disk CMD that there is no evidence for a thin disk
population in this field; instead, the population appears to be
dominated by a thick disk and spheroid.  Thus, it is much more
plausible that the narrow velocity structure is the thick disk and the
broad velocity structure is the spheroid.  These designations would
also explain why the narrow component is significantly rotating but
the broad component is not.

Compared to the disk field, the spheroid field is twice as far from
the galactic center in the plane of the disk, but half the distance
from the galactic center in the plane of the sky.  So, moving our
attention from the disk field to the spheroid field, we expect the
contribution from the disk to decline and the contribution from the
spheroid to increase.  With an exponential disk scale length of
$\approx$5~kpc (Walterbos \& Kennicutt 1988), the disk contribution
must drop from the $\sim$2/3 in the disk field to $<1$\% in the
spheroid field.  Indeed, Figure~\ref{cmdsvels}a shows no indication of
a single narrow component at the Andromeda systemic velocity, as one
would expect if the disk were dominating this position, 51 kpc on the
minor axis.  Furthermore, it is worth noting that the hydrogen column
density in the spheroid field is nearly 25 times smaller than that in the disk
field (Table~\ref{fieldtab}).

Ibata et al.\ (2005) found stars moving with disk velocities at
distances of 15--40 kpc from the galactic center, but they note that
our spheroid field lies beyond the break in the density profile of
their ``disk-like structure.'' They show no evidence that this
structure should comprise a significant population in our field.  The
velocity dispersion in their extended disk is 30 km s$^{-1}$, which is
much narrower than the 80 km s$^{-1}$ we see in our spheroid field.
The velocity dispersion in our spheroid field is in agreement with the
kinematics of the planetary nebulae (Halliday et al.\ 2006;
Hurley-Keller et al.\ 2004), which show a distribution of similar
breadth and evidence for some rotational support.

An additional piece of evidence comes from the similarity of the
stream and spheroid populations, given that the Worthey et al.\ (2005)
hypothesis rests largely on metallicity.  If metallicity alone were
enough to prove that a field in Andromeda is dominated by disk stars,
one could try to argue that our stream field was dominated by disk
stars, too.  However, it is clear from the morphology, HB luminosity,
and kinematics in our stream field that $\sim$75\% of the population
in this field is comprised of two kinematically-cold components
falling toward Andromeda (Kalirai et al.\ 2006b).  There is no way that
the stream is composed of stars residing in the Andromeda disk.

On all of these grounds, one can see that the spheroid field must have
a negligible contribution from stars currently moving in the Andromeda
disk.  It is also clear that the spheroid velocity distribution is not
as hot as one would expect for a hot halo, nor does it reflect the
kinematics of the halo globular cluster system ($\sigma \sim 150$ km
s$^{-1}$; Perrett et al.\ 2002).  The high metallicity and wide age
distribution of the spheroid is likely due to the merger history of
Andromeda, with the spheroid polluted by a combination of disrupted
satellites, stars born in the merger(s), and stars disrupted from the
Andromeda disk.

\section{Summary}
\label{secsumm}

Using deep {\it HST} observations of Andromeda, we have reconstructed
the complete star formation history in three fields: the spheroid,
tidal stream, and outer disk.

In the best-fit model to the spheroid, 40\% of the stars are
metal-rich and younger than 10~Gyr, in stark contrast to our own
Galactic halo.  The data cannot be reproduced by a population of old
stars alone (age~$>$~10~Gyr).  Although the fit is dominated by old
metal-poor stars and young metal-rich stars, a non-negligible
population of young metal-poor stars is also present, implying that at
least some stars in the spheroid were accreted from dwarf galaxies or
formed from relatively pristine infalling material.  Since the
discovery of a metal-rich intermediate-age population in our spheroid
field (Brown et al.\ 2003), various explanations have been put forth
in the literature, including the hypothesis that the disk dominates
all inner spheroid fields (Worthey et al.\ 2005), and the idea that
our spheroid field is contaminated by the tidal stream and not
representative of the inner spheroid in general (Ibata et al.\ 2004).
In the former scenario, the spheroid field is not special, but it is
actually the disk instead of the spheroid, whereas in the latter
scenario, the field is special, because it is the stream and not the
spheroid.  The constraints provided by the population and kinematic
data argue that the spheroid field does not have a significant
contribution from stars currently residing in Andromeda's disk, but
the young metal-rich population may be the result of stars disrupted
from Andromeda's disk by an earlier merger event.  The star count maps
and kinematic data show no evidence for a dominant stream passing
through the spheroid field, as required to explain the similarity
between the spheroid and stream populations by some chance
intersection of the spheroid field with the stream's orbit.
Furthermore, the metallicity distribution in the spheroid field looks
much like that observed in various other fields throughout the inner
spheroid (Ferguson et al.\ 2002; Durrell et al.\ 1994, 2001, 2004).
It is much more likely that the metal-rich populations throughout the
inner spheroid are of intermediate age, as found in our spheroid
field, instead of invoking the pathological situation where these
metal-rich populations are ancient everywhere except in our spheroid
field.

In the best-fit model to the stream, 70\% of the stars are younger
than 10~Gyr.  A detailed comparison of the age and metallicity
distributions in the stream and spheroid shows them to be remarkably
similar but distinct.  It is unclear if the similarity implies that
the stream's progenitor is representative of the objects that formed
the inner spheroid or if the entire inner spheroid is polluted by
stars stripped from the stream's progenitor during its particular
disruption.  The distinction between the disk and stream populations
-- with the stream including old metal-poor stars that are lacking in
the disk -- suggests that the stream is not comprised of stars
disrupted from the Andromeda disk.

The outer disk of Andromeda more closely resembles the thick disk of
the solar neighborhood than either the spheroid or the stream.
Although a trace population of 0.2--1.0~Gyr stars is present, there
are few stars younger than 4~Gyr, and thus the outer disk does not
appear to host a significant thin disk component.  In the best-fit
model to the disk data, 80\% of the stars are younger than 10~Gyr;
indeed, we also showed that these data are consistent with a
population that is completely devoid of stars older than 10~Gyr.  The
minority population of old metal-poor stars in the disk field is
consistent with the field's kinematics, which show a $\sim$33\%
contribution from the spheroid.  If the population in this spheroid
contribution is assumed to be the same as that in our spheroid field,
the resulting model reproduces the data extremely well, and implies
that $\sim$70\% of the stars in the outer disk are 4--8~Gyr old.  The
disk of Andromeda clearly shares the ``G dwarf problem'' seen in the
solar neighborhood.

In the upcoming {\it HST} observing cycle, we will be observing four
more deep fields in the Andromeda spheroid.  One will be at
$\sim$22~kpc on the minor axis, and the other three will be in the
vicinity of $\sim$35~kpc on the minor axis, thus bracketing that point
in the spheroid where there is a transition from a bulge-like
population to one that more closely resembles a canonical halo.  The
star formation history in these additional fields should help to further
disentangle the complex formation history of the Andromeda system
and its various substructures.

\acknowledgements

Support for proposals 9453 and 10265 was provided by NASA through a
grant from STScI, which is operated by AURA, Inc., under NASA contract
NAS 5-26555.  P.G. would like to acknowledge partial support from NSF
grants AST-0307966 and AST-0507483 and NASA/STScI grants GO-10265
and GO-10134.  R.M.R. also acknowledges support from NSF grant AST-0307931
and from NASA/STScI grants GO-9453 and GO-10265.  We are
grateful to P.\ Stetson for providing his DAOPHOT code, and to J.\
Harris for providing his StarFish code.  During our observation
planning, A. Ferguson kindly provided ground images of our fields;
we also thank her for providing the star count
map used in Figure~\ref{mapfig}.  We wish to acknowledge the
assistance D. VandenBerg provided in determining the transformation of
his isochrones to the ACS bandpasses.  D. Taylor, P. Royle, and
D. Soderblom were enormously helpful during the scheduling and
execution of these large {\it HST} programs.  D. Thilker kindly
provided $N_{HI}$ values at our field locations using his published
and unpublished maps of M31.  We thank J. Kalirai and D. Reitzel for
providing the velocity histograms in Figure~\ref{cmdsvels}, and
F. Hammer, A. Font, and M. Fardal for enlightening discussions.


\end{document}